\documentclass{aa}
\usepackage{graphicx}
\usepackage{txfonts}
\usepackage{xcolor}
\begin{document}

   \title{Toward the limits of complexity of interstellar chemistry: Rotational spectroscopy and astronomical search for \textit{n}- and \textit{i}-butanal\thanks{Transition frequencies from this work
   are given for each conformer as supplementary material.
   The data are available at CDS via anonymous
   ftp to cdsarc.u-strasbg.fr (130.79.128.5) or via
   http://cdsweb.u-strasbg.fr/cgi-bin/qcat?J/A+A/}}
   \titlerunning{Rotational spectroscopy and astronomical search for \textit{n}- and \textit{i}-butanal}
   \authorrunning{Sanz-Novo et al.} 

 
   \author{M. Sanz-Novo
          \inst{1,2,3}, A. Belloche \inst{3}, V. M. Rivilla \inst{4},  R. T. Garrod \inst{5}, J. L. Alonso
          \inst{1}, P. Redondo \inst{2}, C. Barrientos \inst{2},  L. Kolesnikov{\'a} \inst{6}, J.C. Valle \inst{1},  L. Rodr{\'i}guez-Almeida \inst{4}, I. Jimenez-Serra \inst{4}, J. Mart{\'i}n-Pintado \inst{4}, H. S. P. M{\"u}ller \inst{7} and K. M. Menten \inst{3}}

   \institute{Grupo de Espectroscop{\'i}a Molecular (GEM), Edificio Quifima, {\'A}rea de Qu{\'i}mica-F{\'i}sica, Laboratorios de Espectroscop{\'i}a y Bioespectroscop{\'i}a, Universidad de Valladolid, Parque Cient{\'i}fico UVa,
Unidad Asociada CSIC, E-47011 Valladolid, Spain.
 jlalonso@qf.uva.es
         \and
           Computational Chemistry Group, Departamento de Qu{\'i}mica F{\'i}sica y Qu{\'i}mica Inorg{\'a}nica, Facultad de Ciencias, Universidad de Valladolid, 47011 Valladolid, Spain.
             carmen.barrientos@uva.es
         \and
           Max-Planck-Institut für Radioastronomie, Auf dem Hügel 69, 53121 Bonn, Germany
         \and
          Centro de Astrobiolog{\'i}a, Consejo Superior de Investigaciones Cient{\'i}ficas–Instituto Nacional de Tecnica Aeroespacial “Esteban Terradas”, 28850 Madrid, Spain
        \and
          Departments of Chemistry and Astronomy, University of Virginia, Charlottesville, VA 22904, USA
         \and
         Department of Analytical Chemistry, University of Chemistry and Technology, Technick{\'a} 5, 166 28 Prague 6, Czech Republic
         \and
          I. Physikalisches Institut, Universit{\"a}t zu K{\"o}ln, Z{\"u}lpicher Str. 77, 50937 K{\"o}ln, Germany
             }

   \date{Received XXXXX XX, XXXX; accepted XXXXX XX, XXXX}

  \abstract
   {In recent times, large organic molecules of exceptional complexity have been found in diverse regions of the interstellar medium.}
   {In this context, we aim to provide accurate frequencies of the ground vibrational state of two key aliphatic aldehydes,  \textit{n}-butanal and its branched-chain isomer, \textit{i}-butanal, to enable their eventual detection in the interstellar medium. We also want to test the level of complexity that interstellar chemistry can reach in regions of star formation.}
   {We employ a frequency modulation millimeter-wave absorption spectrometer to measure the rotational features of \textit{n}- and \textit{i}-butanal. We analyze the assigned rotational transitions of each rotamer separately using the \textit{A}-reduced semirigid-rotor Hamiltonian.
   We use the spectral line survey ReMoCA  performed with the Atacama Large Millimeter/submillimeter Array to search for \textit{n}- and \textit{i}-butanal toward the star-forming region Sgr~B2(N). We also search for both aldehydes toward the molecular cloud G+0.693-0.027 with IRAM 30\,m and Yebes 40\,m observations.
   {The observational results are compared with computational results from a recent gas-grain astrochemical model.}
   }
   {
   {Several thousand rotational transitions belonging to the lowest-energy conformers of two distinct linear and branched isomers have been assigned in the laboratory spectra up to 325 GHz}. 
   A precise set of the relevant rotational spectroscopic constants  has been determined for each structure as a first step toward identifying both molecules in the interstellar medium. 
   {We report non-detections of \textit{n}- and \textit{i}-butanal toward both sources, Sgr~B2(N1S) and G+0.693-0.027. We find that \textit{n}- and \textit{i}-butanal are at least 2--6 and 6--18 times less abundant than acetaldehyde toward Sgr~B2(N1S), respectively, and that \textit{n}-butanal is at least 63 times less abundant than acetaldehyde toward G+0.693-0.027. While propanal is not detected toward Sgr~B2(N1S) either, with an abundance at least 5--11 lower than that of acetaldehyde, propanal is found to be 7 times less abundant than acetaldehyde in G+0.693-0.027.}
   {Comparison with astrochemical models indicates good agreement between observed and simulated abundances (where available). Grain-surface chemistry appears sufficient to reproduce aldehyde ratios in G+0.693-0.027; gas-phase production may play a more active role in Sgr B2(N1S). Model estimates for the larger aldehydes indicate that the observed upper limits may be close to the underlying values.}
   }
   {Our astronomical results indicate that the family of interstellar aldehydes in the Galactic center region is characterized by a drop of one order of magnitude in abundance at each incrementation in the level of molecular complexity.}

   \keywords{ molecular data  --
             ISM:molecules  --
             astrochemistry  --
             line: identiﬁcation -- ISM: individual objects: Sagittarius B2, G+0.693-0.027}

   \maketitle

%

\section{Introduction}
\label{s:Intro}
Our understanding of the chemistry of star-forming regions is steadily growing thanks to advances in laboratory astrochemistry and the construction of radio telescopes with unparalleled regimes of sensitivity. In current astrochemical research, significant endeavors have been made to study the so-called complex organic molecules (COMs; \citealt{Jorgensen20}), which are carbon-based molecules that contain more than six atoms (see \citealt{McGuire18} for a census). Among them, aldehydes are some of the most widespread species in nature, being precursors of many biologically relevant molecules. Fifty years ago, \citet{1969Snyder} reported the detection of formaldehyde (H$_2$CO), the first interstellar aldehyde, in diverse galactic and extragalactic radio sources using the NRAO 140 ft telescope. Since then, several CHO-bearing species with increasing levels of complexity have been discovered, mainly toward the giant molecular cloud complex Sagittarius B2 (Sgr B2), notably acetaldehyde (CH$_3$CHO; \citealt{1973Gottlieb}), glycolaldehyde (CH$_2$(OH)CHO; \citealt{2000Hollis}), which is the simplest sugar-related molecule, propanal (CH$_3$CH$_2$CHO; \citealt{2004Hollis}), and cyanoformaldehyde (CNCHO; \citealt{2008Remijan}).  More complex aldehydes such as lactaldehyde (CH$_3$CH(OH)CH(O); \citealt{2019Alonso}), which is the rational step up in complexity from glycolaldehyde, are yet to be detected.

Astronomical discoveries of new molecular systems have typically been the driving motive in the selection of novel targets to study. Of paramount relevance was the radio-astronomical discovery of isopropyl cyanide, or isobutyronitrile (\textit{i}-C$_3$H$_7$CN, (CH$_3$)$_2$CHCN), the first aliphatic branched-chain molecule ever detected in the interstellar medium \citep[ISM;][]{Belloche14}. Recently, \citet{Kolesnikova17a} performed a thorough analysis of its spectra in the millimeter- and submillimeter-wave regions. Furthermore, the detection of its linear analog, \textit{n}-butyronitrile (\textit{n}-C$_3$H$_7$CN, CH$_3$CH$_2$CH$_2$CN), has been reported by \citet{2009Belloche} in IRAM 30 m observations of Sgr B2. The identification of new branched-chain species in the ISM will unveil the link between the molecular inventory of the Milky Way and the chemical composition of small celestial objects such as comets \citep{Altwegg16} and asteroids, as well as meteorites, their rocky remnants that land on Earth \citep[][]{Pizzarello10,Burton12}, in which several branched-chain amino acids have been found \citep{2017Koga}. In this context, we propose the study of two key aliphatic aldehydes of extraordinary complexity, \textit{normal}-butanal (\textit{n}-C$_4$H$_8$O; CH$_3$CH$_2$CH$_2$CHO) and the branched \textit{iso}-butanal (\textit{i}-C$_4$H$_8$O; (CH$_3$)$_2$CHCHO), which have already been identified in several chondritic meteorites \citep{2019Aponte}, as targets for interstellar detection.

Further increasing the interest of \textit{n}- and \textit{i}-butanal (also known as \textit{n}- and \textit{i}-butyraldehyde and hereafter \textit{n}-PrCHO and \textit{i}-PrCHO, respectively) as astronomical candidates, \citet{2018Abp} revealed that both isomers -- together with distinct C$_3$H$_8$O species, such as propanol and methyl ethyl ether (the latter was already detected in the ISM toward Orion KL by \citealt{Tercero18} and toward IRAS 16293-2422A by \citealt{Manigand2020}) -- could be generated by exposing interstellar analog ices to ionizing radiation. Therefore, they should be detectable in the gas phase after desorption from the icy grains in regions of star formation \citep{2019Abp}. This further suggests that both aldehydes are expected to be formed in the ISM and motivates the astrophysical community even more to search for them. 

However, the lack of accurate millimeter-wave data of \textit{n}- and \textit{i}-PrCHO (see Sect.~\ref{s:Meth}) thus prompted new laboratory spectroscopic measurements over the frequency range from 75 to 325 GHz. In this work, we aim to provide an extensive rotational characterization of both aldehydes in the millimeter-wave region. The analysis of the rotational spectra at these wavelengths will enable us to confidently search for both species toward different regions of the ISM.

Moreover, with the development of new and higher-sensitivity instrumentation, we expect the detection of large molecular systems (i.e., those containing large alkyl chains, but not only) to grow even beyond \textit{i}-propyl cyanide \citep{Belloche14}. This assumption is strongly supported by some of the latest astronomical detections, which include benzonitrile (c-C$_6$H$_5$CN; \citealt{2018McGuire}), cyanocyclopentadiene, its five-membered ring analog \citep{2021McCarthy}, ethanolamine (HOCH$_2$CH$_2$NH$_2$; \citealt{rivilla2021discovery}), and ethyl isocyanate (C$_2$H$_5$NCO; \citealt{rodriguez2021b}), as well as the 1- and 2- isomers of cyanonaphthalene \citep{2021McGuire} and indene \citep[][]{2021Cernicharo,2021Burk}. However, a thoroughly harmonized observational and laboratory effort will be essential to unveil the pathways of chemical evolution of more complex systems as well as to conclusively disclose the ratio between aliphatic and cyclic molecules in the ISM \citep[][]{2016Loomis,2021McCarthyMcG}).

We describe in Sect.~\ref{s:Meth} the experimental setup that was used to measure the spectra of both aldehydes. In Sect.~\ref{s:CQC} we show the complementary quantum chemical computations. Section~\ref{s:Results} presents the results and discussion of the analysis of the millimeter-wave spectra. The astronomical results of a search for \textit{n}- and \textit{i}-PrCHO toward two prominent sources in the Galactic Center, the high-mass star-forming region Sgr B2(N) and the molecular cloud G+0.693-0.027, are reported in Sects.~\ref{s:astro} and~\ref{s:AQRG}. The astronomical results are discussed in a broader astrochemical framework in Sect.~\ref{s:discussion}. Finally, we present in Sect.~\ref{s:conclusions} the conclusions of this work.

\section{Measurement of the rotational spectra}
\label{s:Meth}

Both \textit{n}- and \textit{i}-PrCHO are colorless liquids with a boiling point of 74.8~$^\circ$C and 63.0~$^\circ$C, respectively, which were purchased from Aldrich and used without further purification. They have been previously investigated in the condensed phases by nuclear magnetic resonance (NMR) and infrared spectroscopies \citep[][]{1970Sabrana,Piart1991,Dwive08}. Also, their conformational panoramas have already been explored in the gas phase using microwave and far-infrared spectroscopy \citep[][]{1986Stiefvatera,1989Durig,2012Hotopp}. However, to our knowledge, their spectra had remained uncharted in the millimeter- and submillimeter-wave regions until now. Accurate spectroscopic data at high frequencies are usually mandatory for the eventual astronomical detection of many interstellar molecules in several spectral line surveys, such as the EMoCA and ReMoCA   surveys conducted with Atacama Large Millimeter/submillimeter Array (ALMA) toward Sgr~B2(N) \citep[][]{Belloche2017,Belloche2019} and the IRAM 30m survey of the G+0.693-0.027 giant molecular cloud \citep{2020Rivilla}.

The room-temperature rotational spectra were recorded separately from 75 to 325 GHz using the millimeter-wave absorption spectrometer at the University of Valladolid (for a detailed description of the instrument, see \citealt[][]{Daly14,Alonso16}). In both experiments, we filled our double-pass free-space glass cell with an optimum gas pressure of about 20 $\mu$bar (no external heating was needed for any of the molecules), which was maintained during the whole course of the experiment. The millimeter-wave radiation was generated by multiplying the fundamental signal of an Agilent E8257D microwave synthesizer (up to 20 GHz) by a set of active and passive amplifier-multiplier chains (multiplication factors of 6, 9, 12, and 18 in this case). Also, the output of the synthesizer was frequency modulated at a frequency of 10.2 kHz, with modulation depths varying between 20 to 35 kHz. After the radiation passed through the cell, the signal was detected using solid-state zero-bias detectors and was sent to a phase-sensitive lock-in amplifier with \textit{2f} demodulation (time constant of 30 ms). This demodulation process leads to a line shape that approximates the second derivative of a Gaussian function. We measured the rotational lines employing an average of two up and down frequency scans, also using a Gaussian profile function within the AABS package \citep{Kisiel05}. The experimental uncertainty of the unblended symmetric lines is estimated to be about 30 kHz.

\section{Computational section}
\label{s:CQC}

\begin{figure}
\centerline{\resizebox{1.0\hsize}{!}{\includegraphics[angle=0]{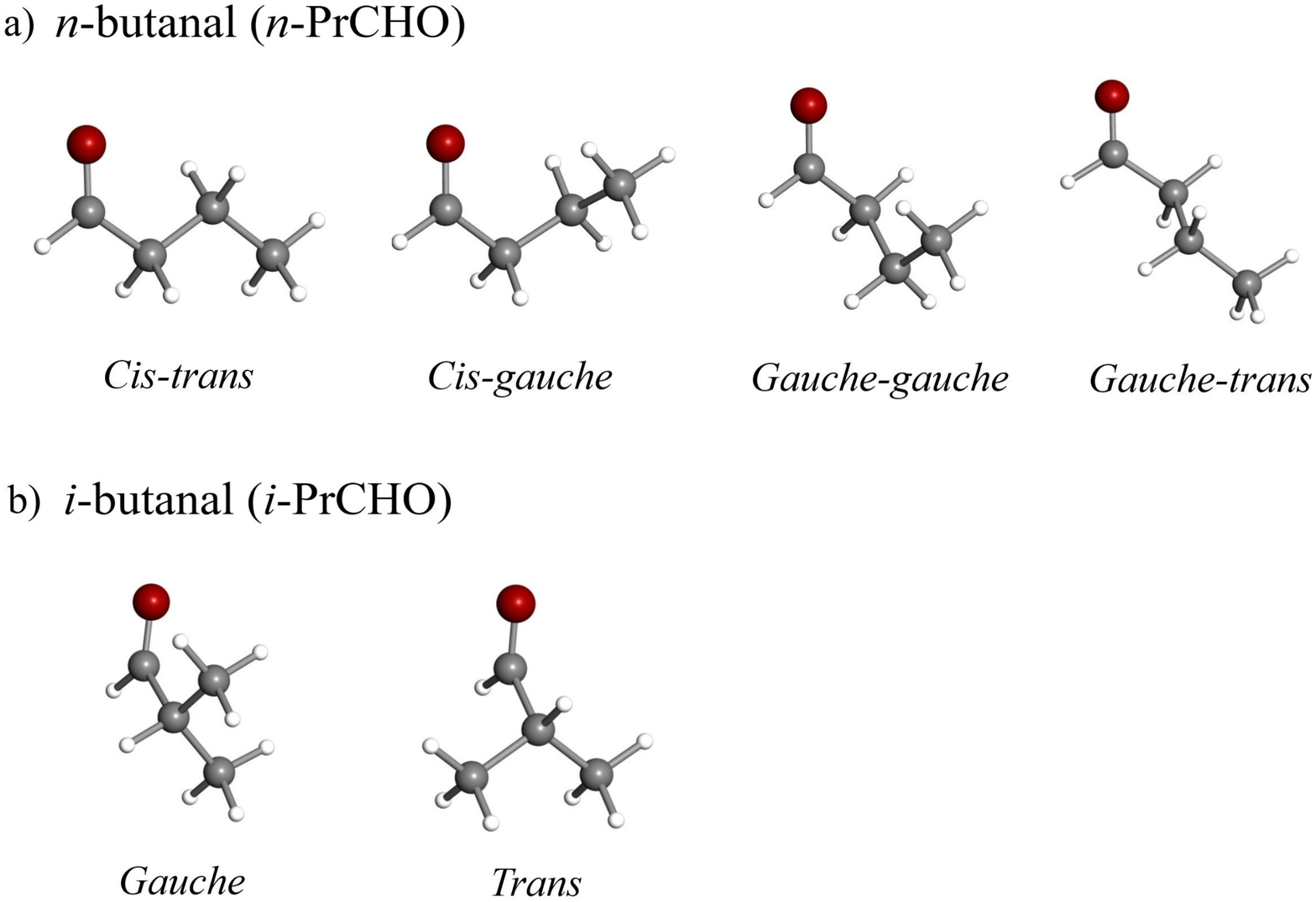}}}
\caption{Structures of the lowest-energy conformations of \textit{n}- and \textit{i}-butanal.}
\label{f:Structures}
\end{figure}

In addition to the previous computational work on \textit{n}-PrCHO  \citep[][]{Klim20,Dwive08,2012Hotopp}, we performed high-level calculations on both \textit{n}- and \textit{i}-PrCHO conformers to accompany the experimental study. The geometries were optimized at the density functional theory level, employing the double-hybrid functional B2PLYP \citep{2006Grimme} combined with Grimme’s D3BJ dispersion \citep[B2PLYPD3;][]{2011Grimme}, in conjunction with the correlation consistent basis set of Dunning aug-cc-pVTZ (correlation-consistent polarized valence triple-zeta including diffuse functions) \citep[][]{89Dunning,Woon93}. Additional optimizations were also carried out using the ab initio coupled-cluster with single and double excitations (CCSD) method \citep{Raghavachari89}. In this case the Dunning’s cc-pVTZ (correlation-consistent polarized valence triple-zeta) basis set was employed. On the optimized geometries, harmonic vibrational frequencies were calculated at both levels of theory. This allows the zero-point vibrational energy (ZPE) correction to be estimated, the nature of the stationary points located on the potential energy surface to be confirmed, and the vibrational contribution to the partition function to be computed. The structures of the low-energy conformers of both isomers are presented in Fig.\ref{f:Structures}, and their harmonic vibrational frequencies calculated at the CCSD/cc-pVTZ level are collected in Table \ref{TableA1}.

To obtain more accurate stability order, single point calculations were carried out on the CCSD/cc-pVTZ optimized geometries. Energies are calculated at the CCSD(T) (coupled-cluster with single and double excitations including triple excitations through a perturbative treatment) level extrapolated to the complete basis set limit, denoted as CCSD(T)/CBS. It is computed from the n$^{-3}$ extrapolation equation \citep{Helgaker97} applied to the CCSD(T) energies calculated with the triple- (n=3 aug-cc-pVTZ) and quadruple-zeta (n=4, aug-cc-pVQZ) basis sets. In Table \ref{t:teortable1} the calculated spectroscopic parameters and relative energies of the conformer of \textit{n}-PrCHO are reported, the corresponding values for the \textit{i}-PrCHO conformers are shown in Table \ref{t:teortable2}. 

To perform the corresponding quantum chemical computations, we used the Gaussian \citep{Frisch16} and CFOUR \citep{Stanton2013} program packages. 

\begin{table*}[!ht]
\begin{center}
\caption{Theoretical ground-state spectroscopic constants for the low-energy \textit{n}-butanal conformers ($A$-Reduction, I$^{r}$-Representation).}
\label{t:teortable1}
\vspace*{0.0ex}
\begin{tabular}{lllll}
\hline\hline
\multicolumn{1}{c}{Parameters} & \multicolumn{1}{c}{\textit{cis-trans}} & \multicolumn{1}{c}{\textit{cis-gauche}} & \multicolumn{1}{c}{\textit{gauche-gauche}} & \multicolumn{1}{c}{\textit{gauche-trans}}\\ 
\hline
A\tablefootmark{(a)}\small (MHz) & 15184.914 &  8535.664 & 9888.586 & 20370.216 \\
B \small (MHz) & 2565.358 & 3610.610 & 3050.824 & 2155.419 \\ 
C \small (MHz) & 2286.380 & 2940.554 & 2610.863 & 2114.933 \\ 
|$\mu$$_a$|, |$\mu$$_b$|, |$\mu$$_c$|\tablefootmark{(b)}\small (D) & 1.5 / 2.0 / 0.0 & 0.7 / 2.3 / 0.7   & 2.3 / 1.5 / 0.7 & 2.6 / 0.4 / 1.2 \\ 
$\Delta$$_J$ \small (kHz) & 0.603 & 3.544 & 6.004 & 0.899 \\ 
$\Delta$$_K$ \small (kHz) & 36.693 & 21.981 & 162.286 & 635.956 \\ 
$\Delta$$_{JK}$ \small (kHz) & -4.179 & -11.516 & -51.531 & -34.504 \\ 
$\delta$$_J$ \small (kHz) & 0.0987 & 1.078 & 1.814 & -0.134 \\ 
$\delta$$_K$ \small (kHz) & -0.0926 & 5.354 & 9.705 & 13.712 \\ 
$V$$_3$\tablefootmark{(c)}  & 1067 & 963  & 997 & 1069 \\ 
$\Delta$E\tablefootmark{(d)} & 0.00 & 0.87 & 3.12 & 3.20 \\ 
\hline 
\end{tabular}
\end{center}
\vspace*{-2.5ex}
\tablefoot {\tablefootmark{(a)}$A$, $B$, and $C$ represent the rotational constants computed at the CCSD/cc-pVTZ level. \tablefootmark{(b)}|$\mu$$_a$|, |$\mu$$_b$|, and |$\mu$$_c$| are the absolute values of the electric dipole moment components (in D). \tablefootmark{(c)} $V$$_3$ is the barrier height to the methyl internal rotation (in cm$^{-1}$).\tablefootmark{(d)} $\Delta$E is the energy calculated at the CCSD(T)/CBS level, taking the ZPE at the CCSD/cc-pVTZ level (in kJ mol$^{-1}$) into account.}
\end{table*}

\begin{table}
\begin{center}
\caption{Theoretical ground-state spectroscopic constants for the two low-energy conformers of \textit{i}-butanal ($A$-Reduction, I$^{r}$-Representation).}
\label{t:teortable2}
\vspace*{0.0ex}
\begin{tabular}{lll}
\hline\hline
\multicolumn{1}{c}{Parameters} & \multicolumn{1}{c}{\textit{gauche}} & \multicolumn{1}{c}{\textit{trans}} \\ 
\hline
A\tablefootmark{(a)}\small (MHz) & 7534.289 & 7757.132 \\
B \small (MHz) & 4125.416 & 3737.809 \\ 
C \small (MHz) & 3000.209 & 2827.664 \\ 
|$\mu$$_a$|, |$\mu$$_b$|, |$\mu$$_c$|\tablefootmark{(b)}\small (D) &  2.4 / 0.8 / 0.9 & 2.7 / 0.0 / 0.7  \\ 
$\Delta$$_J$ \small (kHz) & 1.856 & 0.703 \\ 
$\Delta$$_{JK}$ \small (kHz) & -0.274 & 51.547 \\ 
$\Delta$$_K$ \small (kHz) & 10.485 & -45.357 \\ 
$\delta$$_J$ \small (kHz) & 0.259 & 0.191 \\ 
$\delta$$_K$ \small (kHz) & 3.764 & 25.128 \\ 
$\Delta$E\tablefootmark{(c)} & 0.00 & 1.86 \\ 
\hline 
\end{tabular}
\end{center}
\vspace*{-2.5ex}
\tablefoot {\tablefootmark{(a)}$A$, $B$, and $C$ represent the rotational constants computed at the CCSD/cc-pVTZ level. \tablefootmark{(b)}|$\mu$$_a$|, |$\mu$$_b$|, and |$\mu$$_c$| are the absolute values of the electric dipole moment components (in D). \tablefootmark{(c)} $\Delta$E is the energy calculated at the CCSD(T)/CBS level, taking the ZPE at the CCSD/cc-pVTZ level (in kJ mol$^{-1}$) into account.}
\end{table}

\section{Rotational spectra analysis and discussion}
\label{s:Results}

\subsection{Millimeter-wave spectra of \textit{n}-PrCHO}
\label{s:MSN}

\begin{figure*}
\centering
\includegraphics[width=16cm]{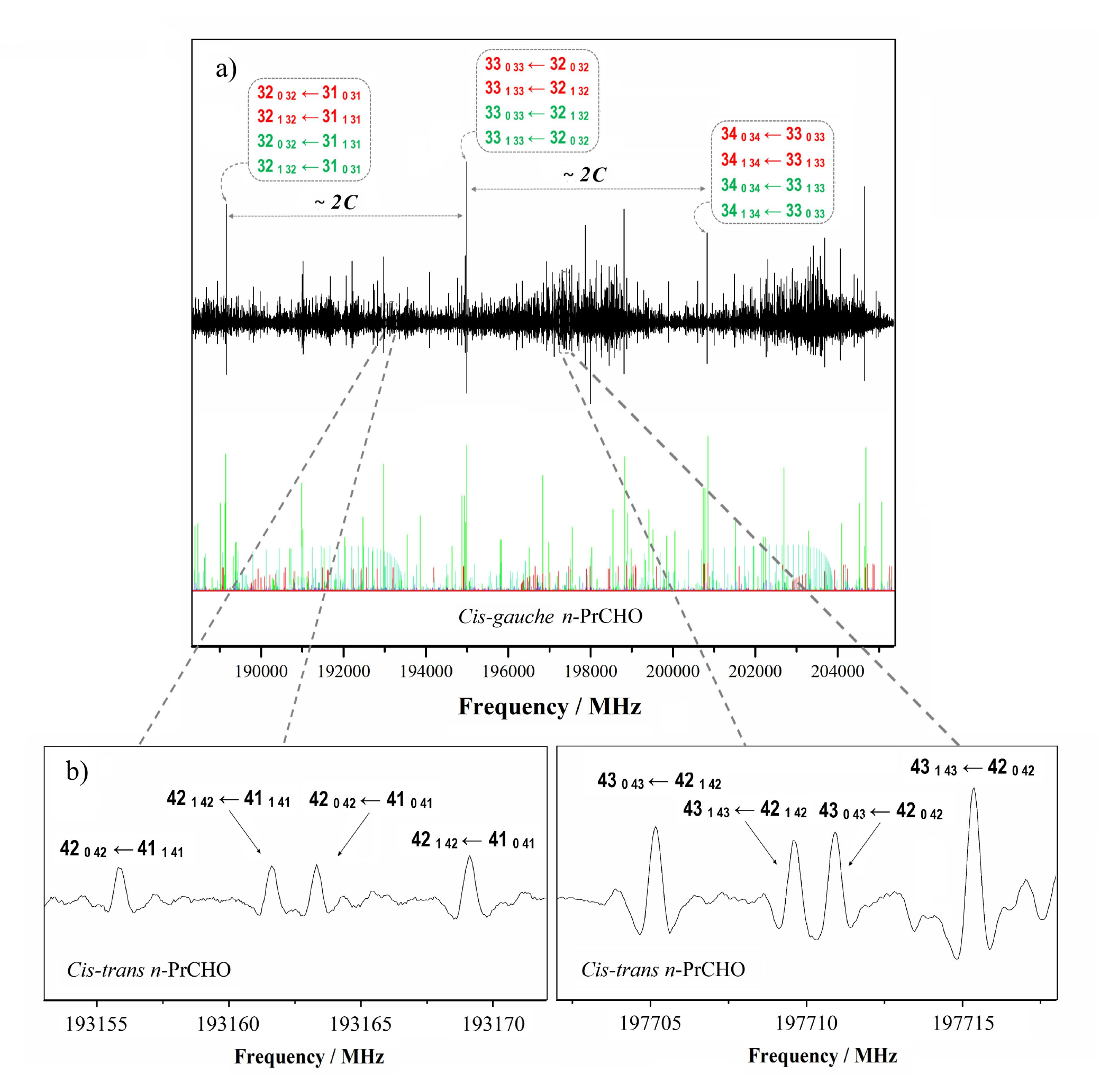}
\caption{Section of the millimeter-wave spectrum of \textit{n}-PrCHO. a) Zoomed view of the millimeter-wave spectrum that shows the dominant quadruple degenerated lines of the \textit{cis-gauche} conformer, which contains pairs of \textit{a}- and \textit{b}-type \textit{R}-branch transitions. The final predicted spectrum of this conformer, computed for 300 K, is given for comparison (\textit{a}-type lines are depicted in red, and \textit{b}-type lines are depicted in green). b) Zoomed-in view of the millimeter-wave spectrum that shows unblended quadruplets of \textit{a}- and \textit{b}-type \textit{R}-branch transitions of the \textit{cis-trans} conformer with \textit{K$_a$} = 0, 1 at \textit{J''} = 41 (\textit{left panel}) and \textit{J''} = 42 (\textit{right panel}). Intensity is given in arbitrary units.}
\label{n-Mmspectrum}
\end{figure*}

In Fig. \ref{n-Mmspectrum}(a) we present a fragment of the room-temperature millimeter-wave spectrum. The spectrum is governed by sets of strong \textit{b}-type \textit{R}-branch (\textit{J}+1) \textit{$_{0}$} $_{\textit{J+1}}$ $\leftarrow$  \textit{J}  \textit{$_{1}$} $_{\textit{J}}$ and (\textit{J}+1)$_{1}$ $_{\textit{J}+1}$ $\leftarrow$  \textit{J} \textit{$_{0}$} $_{\textit{J}}$ transitions, which are separated by about 6 GHz and outshine the rest of the spectral features. Starting from the previous microwave assignments \citep{2012Hotopp}, we found that this patterns approximately coincides with the 2C predicted value of the \textit{cis-gauche} conformer. We noted that at these high frequencies the rotational energy levels corresponding to the lowest \textit{K$_a$} values become quasi degenerate (depicted in \ref{n-Mmspectrum}(a) for \textit{K$_a$} = 0, 1). Thus, a quadruple-degenerated blended line arises from the coalescence of pairs of \textit{b}-type rotational transitions with the corresponding \textit{a}-type lines that involve these energy levels. The early assignments were further extended to other weaker \textit{b}-type transitions. Subsequently, we assigned higher \textit{K$_a$} \textit{R}-branch \textit{a}- and \textit{b}-type transitions together with several \textit{Q}-branch transitions, spanning \textit{J} and \textit{K$_a$} values up to 70 and 31, respectively.
 
In a second step, a closer inspection of the millimeter-wave spectrum was directed to search for the \textit{cis-trans} conformer, which shows a practically prolate behavior. Guided by the previous experimental data \citep{2012Hotopp}, we managed to identify several progressions of \textit{a}- and \textit{b}- type \textit{R}-branch transitions belonging to \textit{cis-trans} \textit{n}-PrCHO, of relatively low intensity (see Fig. \ref{n-Mmspectrum}(b)). Afterward, new rotational lines were assigned and after iterative fittings and predictions, we completed the rotational quantum number’s coverage up to \textit{J} = 69 and \textit{K$_a$} = 20. 

Concurrent to the laboratory work, the barriers of the methyl internal rotation were predicted by high-level B2PLYPD3 quantum chemical calculations to be 11.5 and 12.5 kJ/mol for the \textit{cis-gauche} and \textit{cis-trans} conformers, respectively. Nevertheless, we employed the XIAM program \citep{1996Hart} to predict the methyl group internal rotation splitting as well as the Loomis–Wood type plot technique \citep[][]{1928Loomis,2012Kisiel} to have a thorough look at the spectra. No splittings due to internal rotation motion could be observed for either of the conformers, which is in perfect agreement with the computed barrier heights.

\begin{table*}[!ht]
\begin{center}
\caption{Experimental ground-state spectroscopic constants of \textit{cis-gauche} and \textit{cis-trans} \textit{n-}PrCHO ($A$-Reduction, I$^{r}$-Representation).}
\label{t:mmwtable1}
\vspace*{0.0ex}
\begin{tabular}{lllll}
\hline\hline
\multicolumn{1}{c}{Parameters} & \multicolumn{1}{c}{\textit{cis-gauche}} & \multicolumn{1}{c}{\textit{cis-gauche} (mw)\tablefootmark{(g)}} & \multicolumn{1}{c}{\textit{cis-trans}}& \multicolumn{1}{c}{\textit{cis-trans} (mw)\tablefootmark{(g)}} \\ 
\hline
A\tablefootmark{(a)}\small (MHz) & 8508.53795 (16)\tablefootmark{(f)} & 8508.527 (3) &  15069.38645 (92)  & 15069.347 (5) \\
B \small (MHz) & 3588.81565 (14) & 3588.809 (1) & 2555.992540 (85) & 2555.983(2)\\ 
C \small (MHz) & 2928.57776 (10) & 2928.5749 (9) &  2278.607628 (88)  & 2278.608(1)\\ 
|$\mu$$_a$|, |$\mu$$_b$|, |$\mu$$_c$|\tablefootmark{(b)}\small (D) & Yes / Yes / Yes & Yes / Yes / No & Yes / Yes / No  & Yes / Yes / No\\ 
$\Delta$$_J$ \small (kHz) & 3.648341 (87) & 3.63 (1) &  0.611752 (30)  & 0.50 (5)\\ 
$\Delta$$_{JK}$ \small (kHz) & -12.199378 (19) & -12.04 (8) & -4.23054 (30)  & -4.3 (3)\\ 
$\Delta$$_K$ \small (kHz) & 23.5595 (33) & 23.3 (1) & 37.8014 (89) & -\\ 
$\delta$$_J$ \small (kHz) & 1.110506 (38) & 1.101 (4) & 0.0994029 (45)  & - \\ 
$\delta$$_K$ \small (kHz) & 5.60931 (81) & 5.3(1) & -0.1183 (10)  & -\\ 
$\Phi$$_J$ \small (mHz) & -12.147 (23) & - & 0.0344(35)  & -\\ 
$\Phi$$_{JK}$ \small (Hz) & 0.12797 (67) & -  & 0.004310 (50) & -\\ 
$\Phi$$_K$ \small (Hz) & 0.71765 (40) & - & 0.492 (27) & - \\ 
$\Phi$$_{KJ}$ \small (Hz) & -0.49693 (20) & - & -0.2641 (11) & -\\ 
$\phi$$_{J}$ \small (mHz) & -5.603 (11) & - & - & -\\ 
$\phi$$_{JK}$ \small (Hz) & -0.06452 (33) & - & - & -\\ 
N\tablefootmark{(c)} & 2839  & 58 & 1375 & 16\\ 
$\sigma$ \tablefootmark{(d)}\small (kHz) & 29 & 15 & 28 & 8 \\ 
$\sigma$$_{w}$ \tablefootmark{(e)}\small & 1.02 & - & 0.95 & -\\ 
\hline 
\end{tabular}
\end{center}
\vspace*{-2.5ex}
\tablefoot {\tablefootmark{(a)}$A$, $B$, and $C$ represent the rotational constants. \tablefootmark{(b)}|$\mu$$_a$|, |$\mu$$_b$|, and |$\mu$$_c$| are the absolute values of the electric dipole moment components (in D). \tablefootmark{(c)} $N$ is the number of transitions. \tablefootmark{(d)} $\sigma$ is the rms deviation of the fit.\tablefootmark{(e)} $\sigma$$_{w}$ is the unit-less (weighted) deviation of the fit. \tablefootmark{(f)} Standard error in parentheses in units of the last digit. Their values are close to the 1$\sigma$ standard uncertainties since the unit-less (weighted) deviation of the fit is very close to 1.0.\tablefootmark{(g)} Previously measured microwave data from \citet{2012Hotopp} included for comparison.}
\end{table*}

All in all, we measured more than 2800 rotational transitions for the \textit{cis-gauche} \textit{n}-PrCHO along with 1375 transitions for the \textit{cis-trans} conformer, including in the data set the microwave lines previously observed by Hotopp et al. (2012). Pickett’s SPFIT/SPCAT program package (Pickett 1991) was used to perform the corresponding fits using a Watson’s \textit{A}-reduced Hamiltonian in the I$^{r}$-representation \citep{Watson77}. In Table~\ref{t:mmwtable1} we present the final set of spectroscopic constants for both conformers, which conclusively enhance the previous microwave experimental data measured in the 7.5--18.5 GHz frequency range \citep{2012Hotopp}. It is worth noticing that the frequency coverage (75 to 325 GHz) extends well beyond the emission maximum at the typical temperatures of hot interstellar clouds, which enables a confident search for such species in the ISM. 

Finally, we present the rotational \textit{(Q$_r$)} and vibrational \textit{(Q$_v$)} partition functions of \textit{cis-gauche} and \textit{cis-trans} \textit{n}-PrCHO separately, which are shown in Tables~\ref{t:pfun1} and \ref{t:pfun2}. We used SPCAT (Pickett 1991) to compute \textit{Q$_r$}´s values from first principles at the typical temperatures as applied in the JPL database \citep{Pickett98}. We estimated the vibrational part, Q$_v$, using a harmonic approximation and Eq.~3.60 of \citet{Gordy70}, where only the vibrational modes below 1000 cm $^{-1}$ were taken into consideration. We obtained the frequencies of the normal modes from CCSD calculations (see Table~\ref{TableA1} of the appendix). The full partition function, \textit{Q$_{tot}$}, is therefore the product of \textit{Q$_r$} and \textit{Q$_v$}. It should be noted that an additional conformational correction factor should be introduced in order to derive a proper estimation of the total column density (upper limit) of the molecule.

\begin{table}
\begin{center}
\caption{Rotational and vibrational partition functions of \textit{cis-gauche} \textit{n}-PrCHO.}
\label{t:pfun1}
\vspace*{0.0ex}
\begin{tabular}{lll}
\hline\hline
\multicolumn{1}{c}{Temperature}{\small (K)} & \multicolumn{1}{c}{Q$_r$\tablefootmark{(a)}} & \multicolumn{1}{c}{Q$_v$\tablefootmark{(b)}} \\ 
\hline
9.38 & 514.3036 &  1.0000 \\
18.75 & 1451.3130 &  1.0003 \\ 
37.50 & 4101.0172 &  1.0206 \\ 
75.00 & 11598.5581 &  1.2410 \\
150.00 & 32831.0285 &  2.6021 \\ 
225.00 & 60373.8787 &  6.0802 \\ 
300.00 & 93047.2540 & 14.1692 \\ 
\hline 
\end{tabular}
\end{center}
\vspace*{-2.5ex}
\tablefoot{\tablefootmark{(a)}Q$_r$ is the rotational partition function. \tablefootmark{(b)}Q$_v$ is the vibrational partition function. The total partition function of the conformer is therefore Q$_r$ $\times$ Q$_v$.}
\end{table}

\begin{table}
\begin{center}
\caption{Rotational and vibrational partition functions of \textit{cis-trans} \textit{n}-PrCHO.}
\label{t:pfun2}
\vspace*{0.0ex}
\begin{tabular}{lll}
\hline\hline
\multicolumn{1}{c}{Temperature}{\small (K)} & \multicolumn{1}{c}{Q$_r$\tablefootmark{(a)}} & \multicolumn{1}{c}{Q$_v$\tablefootmark{(b)}} \\ 
\hline
9.38 & 518.5813 &  1.0000 \\
18.75 & 1464.0056 &  1.0020 \\ 
37.50 & 4137.2687 &  1.0486 \\ 
75.00 & 11698.8301 &  1.3624 \\
150.00 & 33095.6561 &  3.1151 \\ 
225.00 & 60821.5883 &  7.6557 \\ 
300.00 & 93661.9204 & 18.3964 \\ 
\hline 
\end{tabular}
\end{center}
\vspace*{-2.5ex}
\tablefoot{\tablefootmark{(a)}Q$_r$ is the rotational contribution to the partition function. \tablefootmark{(b)}Q$_v$ is the vibrational contribution to the function. The total partition function of the conformer is Q$_r$ $\times$ Q$_v$.}
\end{table}

\subsection{Millimeter-wave spectra of \textit{i}-PrCHO}
\label{s:MSI}

\begin{figure}
\centerline{\resizebox{1.00\hsize}{!}{\includegraphics[angle=0]{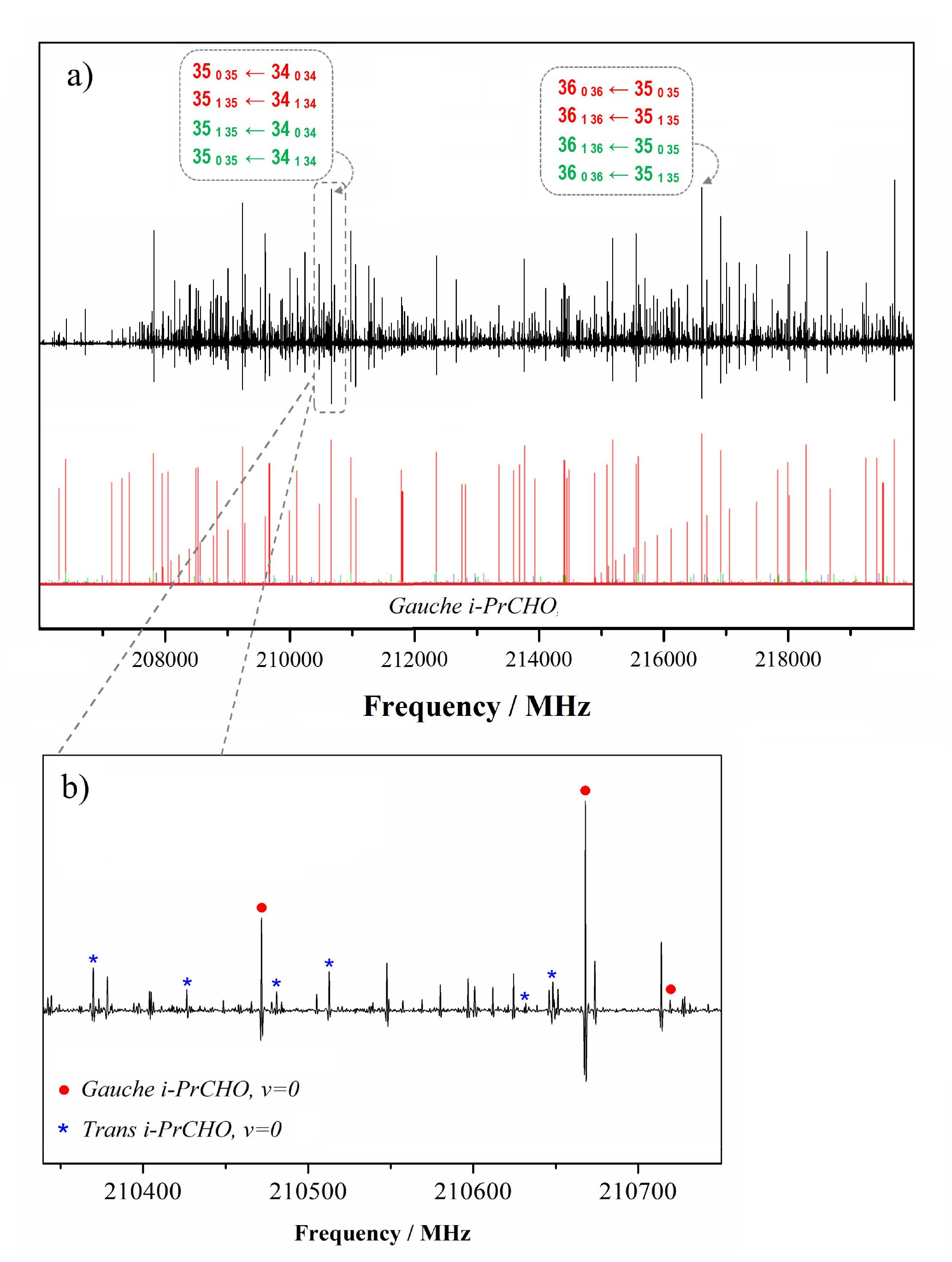}}}
\caption{Room-temperature millimeter-wave spectrum of \textit{i}-PrCHO. a) Portion of the millimeter-wave spectrum of \textit{i}-PrCHO, showing intense \textit{a}-type \textit{R}-branch transitions of the \textit{gauche} conformer. The final computed spectrum for this conformer, computed for 300 K, is given for comparison (\textit{a}-type lines are depicted in red, \textit{b}-type lines are depicted in green, and \textit{c}-type lines are depicted in purple). b) Zoomed-in view of the spectrum, showing several \textit{a}-type \textit{R}-branch lines belonging to \textit{trans} \textit{i}-PrCHO (in blue) together with some intense lines of \textit{gauche} \textit{i}-PrCHO (in red). Intensity is given in arbitrary units.}
\label{f:i-Mmspectrum}
\end{figure}

At the first stage of the spectral analysis, we employed predictions based on the spectroscopic parameters from \citet{1986Stiefvatera}, including the experimental values of the electric dipole-moment components. Given that the two most stable rotameric species of \textit{i}-PrCHO are near prolate asymmetric tops with nonzero $\mu$$_a$ dipole-moment component (\textit{gauche} conformer: |$\mu$$_a$| = 2.43 (2), |$\mu$$_b$| = 0.80 (3), |$\mu$$_c$| = 0.83 (2) Debyes, and \textit{trans} conformer: |$\mu$$_a$| = 2.82 (2), |$\mu$$_b$| = 0.0 by \textit{C$_s$} symmetry, |$\mu$$_c$| = 0.46 (3) Debyes), we first explored the rich millimeter-wave spectrum to search for intense sets of \textit{a}-type \textit{R}-branch transitions. Also in this case, as the quantum number \textit{J} increases, the rotational energy levels that involve the lowest \textit{K$_a$} quantum numbers become closer, and pairs of \textit{a}- and \textit{b}-type lines form quartets, which are already blended into one quadruple-degenerated line at the high frequencies shown in Fig. \ref{f:i-Mmspectrum}(a). After we assigned low \textit{K$_a$} \textit{a}-type transitions, we were able to locate many \textit{b}-type \textit{R}- and \textit{Q}-branch lines, despite the relatively low value of the $\mu$$_b$ dipole moment component. Then, several \textit{c}-type lines were successfully identified as well. This analysis was further extended to higher \textit{K$_a$} \textit{R}- and \textit{Q}-branch transitions. 

Once we finished the analysis of \textit{gauche} \textit{i}-PrCHO, we searched for rotational features attributable to the \textit{trans} conformer. Guided by the previous experimental data (Stiefvater 1986), we managed to identify a much weaker progression of \textit{a}-type \textit{R}-branch transitions belonging to \textit{trans} \textit{i}-PrCHO (see Fig. \ref{f:i-Mmspectrum}(b)). Also, keeping in mind that this conformation belongs to the \textit{C$_s$} molecular symmetry group, a \textit{b}-type spectrum was neither expected nor observed for \textit{trans} \textit{i}-PrCHO. Afterward, we completed the analysis up to \textit{K$_a$} = 12. Also, \textit{c}-type lines were predicted but not observed.

\begin{table*}[!ht]
\begin{center}
\caption{Experimental ground-state spectroscopic constants of \textit{gauche} and \textit{trans} \textit{i-}PrCHO ($A$-Reduction, I$^{r}$-Representation).}
\label{t:mmwtable2}
\vspace*{0.0ex}
\begin{tabular}{lllll}
\hline\hline
\multicolumn{1}{c}{Parameters} & \multicolumn{1}{c}{\textit{gauche}} & \multicolumn{1}{c}{\textit{gauche (mw)}}\tablefootmark{(g)} & \multicolumn{1}{c}{\textit{trans}} & \multicolumn{1}{c}{\textit{trans (mw)}\tablefootmark{(g)}} \\ 
\hline
A\tablefootmark{(a)}\small (MHz) & 7494.61375 (19) \tablefootmark{(f)} & 7494.62 (2) &  7707.8665 (48) & 7707.84 (3)\\
B \small (MHz) & 4107.511659 (79) & 4107.51 (2) & 3736.66463 (78) & 3736.63 (2)\\ 
C \small (MHz) & 2980.699618 (78) & 2980.70 (2) &  2815.08351 (21) & 2815.08 (2)\\ 
|$\mu$$_a$|, |$\mu$$_b$|, |$\mu$$_c$|\tablefootmark{(b)}\small (D) & Yes / Yes / Yes & 2.43 (2) / 0.80 (3) 0.83 (2) & Yes / No / No & 2.82 (2) / 0.0 / 0.46 (3)\\ 
$\Delta$$_J$ \small (kHz) & 1.96051 (56) & 1.90 (45) & 0.82702 (76) & 0.67 (40)\\ 
$\Delta$$_{JK}$ \small (kHz) & -0.29850 (19) & -0.25 (23) & 54.5353 (95) & 48.37 (67)\\ 
$\Delta$$_K$ \small (kHz) & 11.39044 (50) & 11.20 (71) & -47.382 (84) & -19.1 (4.3)\\ 
$\delta$$_J$ \small (kHz) & 0.279096 (22) & 0.28 (2) & 0.24937 (40) & 0.18 (3)\\ 
$\delta$$_K$ \small (kHz) & 3.70440 (20) & 3.63 (19) & 27.0688 (53) & 25.00 (46)\\ 
$\Phi$$_J$ \small (mHz) & 5.472 (14) & - & -1.84 (26) & - \\ 
$\Phi$$_{JK}$ \small (Hz) & 0.062991 (94) & - & 0.01069 (34) & -\\ 
$\Phi$$_K$ \small (Hz) & 0.34139 (58) & - & 45.61 (10) & - \\ 
$\Phi$$_{KJ}$ \small (Hz) & -0.39979 (32) & - & -4.604 (29) & -\\ 
$\phi$$_{J}$ \small (mHz) & 0.1924 (53) & - & -0.883 (135) & - \\ 
$\phi$$_{JK}$ \small (Hz) & -0.047982 (75) & - & 0.0341 (21) & - \\ 
$\phi$$_{K}$ \small (Hz) & -0.093282 (27) & - & - & -\\
N\tablefootmark{(c)} & 3411 & 85 & 513 & 38\\ 
$\sigma$ \tablefootmark{(d)}\small (kHz) & 25 & 76 & 29 & 97\\ 
$\sigma$$_{w}$ \tablefootmark{(e)}\small & 0.79 & - & 0.91 & - \\
\hline 
\end{tabular}
\end{center}
\vspace*{-2.5ex}
\tablefoot {\tablefootmark{(a)}$A$, $B$, and $C$ represent the rotational constants. \tablefootmark{(b)}|$\mu$$_a$|, |$\mu$$_b$|, and |$\mu$$_c$| are the absolute values of the electric dipole moment components (in D). \tablefootmark{(c)} $N$ is the number of transitions. We note that a set of transitions from the previous microwave data from \citet{1986Stiefvatera} was also included in the fit with an adequate weight. \tablefootmark{(d)} $\sigma$ is the rms deviation of the fit.  \tablefootmark{(e)} $\sigma$$_{w}$ is the unit-less (weighted) deviation of the fit. \tablefootmark{(f)} Standard error in parentheses in units of the last digit.\tablefootmark{(g)} Previously measured microwave data from \citet{1986Stiefvatera} included for comparison.}
\end{table*}

In total, almost 3400 rotational transitions were measured for the \textit{gauche} form of \textit{i}-PrCHO up to \textit{J} = 63 and \textit{K$_a$} = 24, while only 500 transitions were recorded for the \textit{trans} form, which is reasonable given that this conformer is higher in energy (see Table \ref{t:teortable2}). The corresponding fits and predictions were made in terms of Watson’s \textit{A}-reduced Hamiltonian in I$^{r}$-representation \citep{Watson77} with the Pickett’s SPFIT/SPCAT program package \citep{Pickett91}. Hence, the present experimental data enabled the determination of accurate rotational constants and the full set of quartic and most of the sextic centrifugal distortion constants, showing a remarkable rms deviation of less than 25 kHz for \textit{gauche} \textit{i}-PrCHO. The final spectroscopic set, which is listed in Table~\ref{t:mmwtable2}, definitively improves the previous microwave experimental data (measured up to 40 GHz) by up to even four orders of magnitude \citep{1986Stiefvatera}. Also, from a comparison between the experimental and theoretically predicted parameters, we can infer that the rotational constants as well as the quartic centrifugal distortion constants are in very good agreement with the CCSD predicted ones. The line catalogs derived from the accurate spectroscopic parameters should be reliable enough for all astronomical observations. The full list of measured transitions in the laboratory for \textit{n}- and \textit{i}-butanal is only available at the CDS.

As reported in Sect. 4.1 for the linear isomer, in Table~\ref{t:pfun3} we provide the values of the rotational \textit{(Q$_r$)} and vibrational \textit{(Q$_v$)} partition functions of \textit{gauche} \textit{i}-PrCHO, which are needed to obtain reliable line intensities. We computed the values of Q$_r$ from first principles up to \textit{J}=150 at different temperatures, using the Pickett's program \citep{Pickett98}.  We performed calculations at the CCSD/cc-pVTZ level of theory to obtain the frequency of the normal modes (see Table A3). Afterward, we predicted the vibrational part, \textit{Q$_v$}, using a harmonic approximation and a simple formula that accords to Eq. (3.60) of \citet{Gordy70}. Finally, we computed the vibrational contribution to the total partition function by taking into account the lowest vibrational modes up to 1000 cm${^{-1}}$. The complete partition function, \textit{Q$_{tot}$}, is then the product of \textit{Q$_r$} and \textit{Q$_r$}. Again, a conformational correction factor is still needed to obtain a proper estimation of the total column density (upper limit) of the molecule.

\begin{table}
\begin{center}
\caption{Rotational and vibrational partition functions of \textit{gauche} \textit{i}-PrCHO.}
\label{t:pfun3}
\vspace*{0.0ex}
\begin{tabular}{lll}
\hline\hline
\multicolumn{1}{c}{Temperature}{\small (K)} & \multicolumn{1}{c}{Q$_r$\tablefootmark{(a)}} & \multicolumn{1}{c}{Q$_v$\tablefootmark{(b)}} \\ 
\hline
9.38 & 507.7789 &  1.0000 \\
18.75 & 1432.7594 &  1.0007 \\ 
37.50 & 4048.2084 &  1.0277 \\ 
75.00 & 11337.6188 &  1.2408 \\
150.00 & 32395.4138 &  2.5019 \\ 
225.00 & 59557.2187 &  5.8979 \\ 
300.00 & 91764.5821 & 14.1391 \\ 
\hline 
\end{tabular}
\end{center}
\vspace*{-2.5ex}
\tablefoot{\tablefootmark{(a)}Q$_r$ is the rotational contribution to the partition function. We took $J$=150 as the maximum value. \tablefootmark{(b)}Q$_v$ is the vibrational partition function. The total partition function of the conformer is the following product: Q$_r$ $\times$ Q$_v$.}
\end{table}

\section{Search for \textit{n-} and \textit{i-}butanal toward Sgr~B2(N1)}
\label{s:astro}

\subsection{Observations}
\label{ss:observations}

The imaging spectral line survey ReMoCA  toward Sgr~B2(N) was carried out with ALMA during its cycle 4. \citet{Belloche19} reported the details about the 
observations and data reduction of the survey. We summarize here the main 
features. The full frequency range from 84.1~GHz to 114.4~GHz was covered 
at a spectral 
resolution of 488~kHz (1.7 to 1.3~km~s$^{-1}$) with five frequency tunings.
The achieved sensitivity per spectral channel varies between 
0.35~mJy~beam$^{-1}$ and 1.1~mJy~beam$^{-1}$ (rms) depending on the setup, with 
a median value of 0.8~mJy~beam$^{-1}$. The phase center is located at the 
equatorial position ($\alpha, \delta$)$_{\rm J2000}$= 
($17^{\rm h}47^{\rm m}19{\fs}87, -28^\circ22'16{\farcs}0$), which is half-way
between the two hot molecular cores Sgr~B2(N1) and Sgr~B2(N2). The angular resolution (half-power beam width, HPBW) ranges from $\sim$0.3$\arcsec$ to $\sim$0.8$\arcsec$ with a 
median value of 0.6$\arcsec$ that corresponds to $\sim$4900~au at the distance of Sgr~B2 \citep[8.2~kpc;][]{Reid19}. We used here an improved version of the 
data reduction, as described in \citet{Melosso20}.

For this work we analyzed the spectrum obtained toward the position 
Sgr~B2(N1S) at ($\alpha, \delta$)$_{\rm J2000}$= 
($17^{\rm h}47^{\rm m}19{\fs}870$, $-28^\circ22\arcmin19{\farcs}48$). This 
position is offset by about 1$\arcsec$ to the south of the main hot core 
Sgr~B2(N1) and was chosen by \citet{Belloche19} because of its lower continuum 
opacity compared to the peak of the hot core. To analyze the spectrum of 
Sgr~B2N1S), we produced synthetic spectra under the assumption of local 
thermodynamic 
equilibrium (LTE) using the software Weeds \citep[][]{Maret11}. The high 
densities of the regions where hot-core emission is detected in Sgr~B2(N)
\citep[$>1 \times 10^{7}$~cm$^{-3}$; see][]{Bonfand19} justify this assumption. 
We derived a best-fit synthetic spectrum for each molecule separately, and 
then added the contributions of all identified molecules together. Each 
species was modeled with a set of five parameters: size of the emitting region 
($\theta_{\rm s}$), column density ($N$), temperature ($T_{\rm rot}$), linewidth 
($\Delta V$), and velocity offset ($V_{\rm off}$) with respect to the assumed 
systemic velocity of the source, $V_{\rm sys}=62$~km~s$^{-1}$.

\subsection{Non-detection of butanal}
\label{ss:nondetection}

In order to search for \textit{n-} and \textit{i-}butanal toward Sgr~B2(N1S),
we relied on the LTE parameters derived for acetaldehyde. The latter is
clearly detected in its torsional ground state and its first two torsionally
excited states (see Figs.~\ref{f:spec_ch3cho_ve0}--\ref{f:spec_ch3cho_ve2}).
We used the spectroscopic predictions of \citet{Smirnov14} for acetaldehyde.
We selected the transitions of this molecule that are not too much contaminated by emission or
absorption from other species to build a population diagram 
(Fig.~\ref{f:popdiag_ch3cho}). Due to the small energy range covered by the 
transitions of the torsional ground state reported in this diagram and the
residual contamination from unidentified species, the rotational temperature 
within the torsional ground state is not constrained. Only a fit taking
into account the three torsional states allowed us to derive a temperature, 
which is found to be high, as reported in Table~\ref{t:popfit}. We assumed that 
the torsional states are populated according to LTE, that is, that the 
torsional temperature is equal to the rotational temperature, and we adopted 
a temperature of 250~K to compute the LTE synthetic spectra shown in red in 
Figs.~\ref{f:spec_ch3cho_ve0}--\ref{f:spec_ch3cho_ve2}. The LTE parameters 
derived for acetaldehyde are reported in Table~\ref{t:coldens}. Given that many lines of acetaldyde are detected with a high signal-to-noise ratio and that assuming LTE does not seem to be a limiting factor in reproducing the line intensities of this molecule, we expect that the uncertainty on its column density lies below 10\%.

\begin{table}
 \begin{center}
 \caption{
 Rotational temperature of acetaldehyde derived from its population diagram toward Sgr~B2(N1S).
}
 \label{t:popfit}
 \vspace*{0.0ex}
 \begin{tabular}{lll}
 \hline\hline
 \multicolumn{1}{c}{Molecule} & \multicolumn{1}{c}{States\tablefootmark{a}} & \multicolumn{1}{c}{$T_{\rm fit}$\tablefootmark{b}} \\ 
  & & \multicolumn{1}{c}{\small (K)} \\ 
 \hline
CH$_3$CHO & $\varv_{
m t}=0$, $\varv_{\rm t}=1$, $\varv_{\rm t}=2$ &   290 (20) \\ 
\hline 
 \end{tabular}
 \end{center}
 \vspace*{-2.5ex}
 \tablefoot{
 \tablefoottext{a}{Vibrational states that were taken into account to fit the population diagram.}
 \tablefoottext{b}{The standard deviation of the fit is given in parentheses. As explained in Sect.~3 of \citet{Belloche16} and in Sect.~4.4 of \citet{Belloche19}, this uncertainty is purely statistical and should be viewed with caution. It may be underestimated.}
 }
 \end{table}

\begin{table*}[!ht]
 \begin{center}
 \caption{
 Parameters of our best-fit LTE model of acetadehyde toward Sgr~B2(N1S) and upper limits for propanal and butanal.
}
 \label{t:coldens}
 \vspace*{-1.2ex}
 \begin{tabular}{lcrcccccccr}
 \hline\hline
 \multicolumn{1}{c}{Molecule} & \multicolumn{1}{c}{Status\tablefootmark{a}} & \multicolumn{1}{c}{$N_{\rm det}$\tablefootmark{b}} & \multicolumn{1}{c}{Size\tablefootmark{c}} & \multicolumn{1}{c}{$T_{\mathrm{rot}}$\tablefootmark{d}} & \multicolumn{1}{c}{$N$\tablefootmark{e}} & \multicolumn{1}{c}{$F_{\rm vib}$\tablefootmark{f}} & \multicolumn{1}{c}{$F_{\rm conf}$\tablefootmark{g}} & \multicolumn{1}{c}{$\Delta V$\tablefootmark{h}} & \multicolumn{1}{c}{$V_{\mathrm{off}}$\tablefootmark{i}} & \multicolumn{1}{c}{$\frac{N_{\rm ref}}{N}$\tablefootmark{j}} \\ 
  & & & \multicolumn{1}{c}{\small ($''$)} & \multicolumn{1}{c}{\small (K)} & \multicolumn{1}{c}{\small (cm$^{-2}$)} & & & \multicolumn{1}{c}{\small (km~s$^{-1}$)} & \multicolumn{1}{c}{\small (km~s$^{-1}$)} & \\ 
 \hline
 CH$_3$CHO, $\varv_{\rm t}=0$$^\star$ & d & 15 &  2.0 &  250 &  6.7 (17) & 1.09 & -- & 5.0 & 0.0 &       1 \\ 
 \hspace*{9.2ex} $\varv_{\rm t}=1$ & d & 11 &  2.0 &  250 &  6.7 (17) & 1.09 & -- & 5.0 & 0.0 &       1 \\ 
 \hspace*{9.2ex} $\varv_{\rm t}=2$ & d & 5 &  2.0 &  250 &  6.7 (17) & 1.09 & -- & 5.0 & 0.0 &       1 \\ 
\hline 
\multicolumn{10}{c}{Upper limits assuming 250~K} \\ 
 \textit{s}-C$_2$H$_5$CHO, $\varv=0$ & n & 0 &  2.0 &  250 & $<$  1.3 (17) & 4.46 & -- & 5.0 & 0.0 & $>$     5.0 \\ 
 \textit{g}-C$_2$H$_5$CHO, $\varv=0$ & n & 0 &  2.0 &  250 & $<$  4.5 (17) & 4.46 & -- & 5.0 & 0.0 & $>$     1.5 \\ 
\hline 
 \textit{ct-n}-C$_3$H$_7$CHO, $\varv=0$ & n & 0 &  2.0 &  250 & $<$  3.4 (17) & 10.3 & 1.66 & 5.0 & 0.0 & $>$     2.0 \\ 
 \textit{cg-n}-C$_3$H$_7$CHO, $\varv=0$ & n & 0 &  2.0 &  250 & $<$  5.1 (17) & 8.11 & 2.52 & 5.0 & 0.0 & $>$     1.3 \\ 
 \textit{g-i}-C$_3$H$_7$CHO, $\varv=0$ & n & 0 &  2.0 &  250 & $<$  1.2 (17) & 7.97 & -- & 5.0 & 0.0 & $>$     5.6 \\ 
\hline 
\multicolumn{10}{c}{Upper limits assuming 170~K} \\ 
 \textit{s}-C$_2$H$_5$CHO, $\varv=0$ & n & 0 &  2.0 &  170 & $<$  6.2 (16) & 2.07 & -- & 5.0 & 0.0 & $>$      11 \\ 
 \textit{g}-C$_2$H$_5$CHO, $\varv=0$ & n & 0 &  2.0 &  170 & $<$  6.2 (17) & 2.07 & -- & 5.0 & 0.0 & $>$     1.1 \\ 
\hline 
 \textit{ct-n}-C$_3$H$_7$CHO, $\varv=0$ & n & 0 &  2.0 &  170 & $<$  1.2 (17) & 3.96 & 1.54 & 5.0 & 0.0 & $>$     5.5 \\ 
 \textit{cg-n}-C$_3$H$_7$CHO, $\varv=0$ & n & 0 &  2.0 &  170 & $<$  2.3 (17) & 3.26 & 2.85 & 5.0 & 0.0 & $>$     2.9 \\ 
 \textit{g-i}-C$_3$H$_7$CHO, $\varv=0$ & n & 0 &  2.0 &  170 & $<$  3.8 (16) & 3.13 & -- & 5.0 & 0.0 & $>$      18 \\ 
\hline 
 \end{tabular}
 \end{center}
 \vspace*{-2.5ex}
 \tablefoot{
 \tablefoottext{a}{d: detection, n: non-detection}
 \tablefoottext{b}{Number of detected lines \citep[conservative estimate; see Sect.~3 of][]{Belloche16}. One line of a given species may mean a group of transitions of that species that are blended together.}
 \tablefoottext{c}{Source diameter (full width at half maximum).}
 \tablefoottext{d}{Rotational temperature.}
 \tablefoottext{e}{Total column density of the molecule. $x$ ($y$) means $x \times 10^y$. An identical value for all listed vibrational states of a molecule means that LTE is an adequate description of the vibrational excitation. For propanal, the two conformers are considered as parts of the same species, and each column density corresponds to the total column density of the molecule. For \textit{n}-butanal, the two conformers were modeled as independent species, and a conformer correction ($F_{\rm conf}$) was applied a posteriori such that each column density corresponds to the total column density of \textit{n}-butanal. For \textit{i}-butanal, the gauche conformer is considered as an independent species, and its column density corresponds to the column density of this conformer only.}
 \tablefoottext{f}{Correction factor that was applied to the column density to account for the contribution of vibrationally excited states in the cases where this contribution was not included in the partition function of the spectroscopic predictions.}
 \tablefoottext{g}{Correction factor that was applied to the column density to account for the contribution of other conformers in the cases where this contribution could be estimated but was not included in the partition function of the spectroscopic predictions.}
 \tablefoottext{h}{Linewidth (full width at half maximum).}
 \tablefoottext{i}{Velocity offset with respect to the assumed systemic velocity of Sgr~B2(N1S), $V_{\mathrm{sys}} = 62$ km~s$^{-1}$.}
 \tablefoottext{j}{Column density ratio, with $N_{\rm ref}$ the column density of the previous reference species marked with a $\star$.}
 }
 \end{table*}


Assuming that the more complex aldehydes propanal and butanal trace the same
region as acetaldehyde, we produced LTE synthetic spectra for these species
adopting the same parameters as acetaldehyde with only the column density left
as a free parameter. We employed the spectroscopic predictions derived for
butanal in Sect.~\ref{s:Results} and the predictions available in the Cologne Database of Molecular Spectroscopy (CDMS) catalog 
\citep[][]{Mueller05} for \textit{syn} and \textit{gauche} propanal (tag 58505 
version 2 and tag 58519 version 1, respectively), which are based on 
\citet{Zingsheim17}. The LTE synthetic spectra of propanal and butanal were 
used to search for emission of these species in the ReMoCA  survey toward 
Sgr~B2(N1S). None of them is detected, as illustrated in 
Figs.~\ref{f:spec_c2h5cho-s_ve0}--\ref{f:spec_ch3-2-chcho-g_ve0}. The upper 
limits on their column densities are reported in
Table~\ref{t:coldens}. Estimating the uncertainties on these upper limits is not straightforward because of the uncertainties affecting the level of the baseline and because of the spectral confusion that leads to blends with identified or unidentified species. These uncertainties could be as high as a factor of two.

Table~\ref{t:coldens} indicates that propanal is at least $\sim$5~times less 
abundant than acetaldehyde in Sgr~B2(N1S). As indicated in Table~\ref{t:coldens}, we get the most stringent upper limit on the column density of \textit{normal}-butanal from its \textit{cis-trans (ct)} conformer. We find that \textit{normal}-butanal is at least 
$\sim$2~times less abundant than acetaldehyde. Finally, we find that 
\textit{iso}-butanal is at least $\sim$6~times less abundant than acetaldehyde.

Afterward, given that our chemical model suggests that propanal and butanal may trace lower temperatures than acetaldehyde (see Sect.~\ref{s:discussion}), we also computed column density upper limits assuming a temperature of 170~K instead of 250~K. This has a significant impact on the upper limits because the vibrational partition functions of propanal and butanal are a steep function of temperature in this temperature range. Under this assumption, Table~\ref{t:coldens} indicates that propanal, \textit{normal}-butanal, and \textit{iso}-butanal are at least 11, 6, and 18 times less abundant than acetaldehyde in Sgr~B2(N1S).

\section{Search for \textit{n-} and \textit{i-}butanal toward the G+0.693-0.027 molecular cloud}
\label{s:AQRG}

\subsection{Observations}

The high-sensitivity spectral survey toward the G+0.693-0.027 molecular cloud was carried out using the Yebes 40m telescope (Guadalajara, Spain) and the IRAM 30m telescope (Granada, Spain). Detailed information of the observational survey is presented in \citet{rivilla2021discovery,rivilla2021hncn} and \citet{rodriguez2021a,rodriguez2021b}. The observations used the position switching mode, and were centered at $\alpha$(J2000.0)=$\,$17$^{\rm h}$47$^{\rm m}$22$^{\rm s}$, $\delta$(J2000.0)=$\,-$28$^{\circ}$21$^{\prime}$27 $^{\prime\prime}$, with the off position at (-885$^{\prime\prime}$,+290$^{\prime\prime}$).

\subsection{Non-detection of butanal}

We implemented the spectroscopy presented in this work into the MADCUBA package{\footnote{Madrid Data Cube Analysis on ImageJ is a software developed at the Center of Astrobiology (CAB) in Madrid; http://cab.inta-csic.es/madcuba/Portada.html.}}
\citep[version 26/07/2021,][]{martin2019}, and used the SLIM (Spectral Line Identification and Modeling) tool to generate synthetic spectra of the different conformers of butanal under the LTE assumption, which were compared with the observed spectrum. To evaluate if the transitions are blended with emission from other species, we also considered the LTE model that predicts the total contribution of the more than 120 species identified so far toward G+0.693-0.027 (e.g., \citealt{requenatorres2008,zeng2018,rivilla2019,rivilla2020,jimenez2020,rodriguez2021a,rodriguez2021b,rivilla2021discovery},b).

None of the butanal conformers were clearly detected in the observed data. In most cases the transitions appear strongly blended with brighter lines of abundant species.
To compute the upper limits of the column density for the different conformers, we searched for the brightest predicted spectral features that are not blended with emission from other molecules. The transitions used for $cg$-$n$-C$_3$H$_7$CHO, $ct$-$n$-C$_3$H$_7$CHO, and $g$-$i$-C$_3$H$_7$CHO are 7$_{5,2}-$6$_{4,3}$ (at 92.905774 GHz), 10$_{0,10}-$9$_{0,9}$ (at 47.6364885 GHz), and 6$_{0,6}-$5$_{0,5}$ (at 38.1517752 GHz), respectively.
Using MADCUBA, we determined the upper limit of the column density using the 3$\sigma$ value of the integrated intensity (see details in \citealt{martin2019}), and the same physical parameters used for $s-$C$_2$H$_5$CHO (Table \ref{tab:g0693}): $T_{\rm rot}$=12~K, V$_{\rm LSR}$=69~km~s$^{-1}$ and $\Delta$V=21 km s$^{-1}$.
Since the excitation temperatures of the molecules in G+0.693 are low ($T_{\rm ex}$=5$-$20~K), we did not use the vibrational contribution of the partition function (Q$_v$) in this case. 
The two conformers of $n-$butanal were modeled as independent species without taking into account their zero-point energy, since the low excitation temperatures in G+0.693 do not allow the relative energies of the conformers to be overcome.  
The derived upper limits for the two conformers of \textit{normal}-butanal and for the $g$ conformer of \textit{iso}-butanal are presented in Table \ref{tab:g0693}. 
The upper limit of the lowest-energy conformer, the $ct$-conformer of \textit{normal}-butanal, is $<$0.8$\times$10$^{13}$ cm$^{-2}$.

As we did for Sgr B2(N1S), we also fitted the acetaldehyde (CH$_3$CHO) emission toward G+0.693. The details of the analysis are presented in Appendix \ref{app-ch3cho}, and the derived parameters are shown in Table \ref{tab:g0693}.

We show in Table \ref{tab:ratios} the molecular ratios of the different aldehydes, using also the LTE results of 
$s-$C$_2$H$_5$CHO from
\citet{rivilla2020}.
The acetaldehyde / $s-$propanal ratio is $\sim$7,  the $s-$propanal / $ct-n-$butanal ratio is $>$9, and the acetaldehyde / $ct-n-$butanal ratio is $>$ 63 . The relative molecular ratios found in G+0.693 indicate that each jump in complexity in the aldehyde family implies an abundance drop of around one order of magnitude, similarly to other chemical families such as alcohols, thiols, and isocyanates (\citealt{rodriguez2021a,rodriguez2021b}).

\begin{table*}
\centering
\caption{Derived physical parameters for acetaldehyde, propanal, and butanal toward the G+0.693-0.027 molecular cloud.}
\begin{tabular}{ c  c c c c c c  }
\hline
\hline
 Molecule & $N$   &  $T_{\rm rot}$ & $V_{\rm LSR}$ & $\Delta V$  & Abundance$^a$ & Ref.$^b$  \\
 & ($\times$10$^{13}$ cm$^{-2}$) & (K) & (km s$^{-1}$) & (km s$^{-1}$) & ($\times$10$^{-10}$) &   \\
\hline
CH$_3$CHO  &  50$\pm$1 & 9.4$\pm$0.1 & 69.5$\pm$0.1 & 21.0$\pm$0.3 & 37 &  (1)  \\  
\hline
$s$-C$_2$H$_5$CHO & 7.4$\pm$1.5 &  12.0$\pm$0.8 & 69 & 21 & 5$\pm$2 &  (2) \\
\hline
$ct$-$n$-C$_3$H$_7$CHO$^c$ & $<$0.8 & 12  & 69 & 21 & $<$0.6 &  (1) \\
$cg$-$n$-C$_3$H$_7$CHO$^c$  & $<$0.5 & 12  & 69 & 21 & $<$0.4 &  (1) \\
$g$-$i$-C$_3$H$_7$CHO$^c$ & $<$1.6 & 12  & 69 & 21& $<$1.2 &  (1) \\
\hline 
\end{tabular}
\label{tab:parameters}
\vspace{0mm}
\vspace*{-1.5ex}
 \tablefoot{
 \tablefoottext{a}{We adopted $N_{\rm H_2}$=1.35$\times$10$^{23}$ cm$^{-2}$ from \citet{martin2008}.}
 \tablefoottext{b}{References: (1) This work; (2) \citet{requenatorres2008,rivilla2020}}
 \tablefoottext{c}{For \textit{n}-butanal, the two conformers were modeled as independent species without applying a conformer correction. For \textit{i}-butanal, the gauche conformer is considered as an independent species, and its column density corresponds to the column density of this conformer.}
 }
\label{tab:g0693}
\end{table*}

\begin{table}
\centering
\tabcolsep 4.5pt
\caption{Molecular ratios of aldehydes toward Sgr~B2(N1S) and G+0.693-0.027.}
\begin{tabular}{ c c c }
\hline
\hline
Ratio & Sgr~B2(N1S)    &  G+0.693-0.027 \\ 

\hline
%
CH$_3$CHO / C$_2$H$_5$CHO$^{(a)}$ & $>$5$-$11 & $\sim$7 \\
C$_2$H$_5$CHO$^{(a)}$ / C$_3$H$_7$CHO$^{(b)}$  & -- & $>$9 \\
CH$_3$CHO / C$_3$H$_7$CHO$^{(b)}$ &  $>$2$-$6 &  $>$63 \\
\hline 
\end{tabular}
\label{tab:parameters}
\vspace{0mm}
{\\ (a) Using the $s-$ conformer; (b) Using the $ct-n$ conformer.}
\label{tab:ratios}
\end{table}

\section{Discussion: Interstellar chemistry of acetaldehyde, propanal, and butanal}
\label{s:discussion}


Of the three aldehydes for which column densities or upper limits are presented in Tables 9 and 10, typical astrochemical kinetics models consider only acetaldehyde. However, a treatment for the interstellar formation and destruction of propanal was included in the network of 
\citet{Garrod13}, 
as part of a broader effort to simulate the chemistry of large organic molecules including glycine. Production of propanal could proceed through the addition of various functional-group radicals on dust-grain surfaces during the star-formation process. The network was developed further in various publications, including 
\citet{Garrod17}, 
who added the butyl cyanides, pentanes, and some associated species. The larger aldehydes were not included in that network, however. The most recent usage of this chemical network was presented by 
\citet{Garrod21}. 
In that work, new treatments of so-called non-diffusive reaction mechanisms on grain surfaces and within the dust-grain ice mantles 
\citep{Jin20} 
were applied to a full hot-core chemical simulation. This more complete treatment of grain chemistry enabled the possible production of complex organics at very cold, early times in the evolution of a hot core, as well as during the warm-up of the dust, and -- where permitted by the viability of appropriate reactions -- in the gas phase, following the thermal desorption of the icy dust-grain mantles.

Despite the absence of butanal from the chemical networks, the results of the \citet{Garrod21} 
{hot-core} astrochemical kinetics models are instructive, especially in comparison with the observational results toward Sgr B2(N1S). The models showed their best agreement with other molecular abundance data from the ReMoCA  survey when a long warm-up timescale (i.e., the {slow} model setup) was used; the same model is compared here with the present data. The peak gas-phase fractional abundance of acetaldehyde with respect to total hydrogen in the model is around $2.5 \times 10^{-7}$, providing a peak ratio with gas-phase methanol of $\sim$1:30. This value coincides exactly with the observational ratio, using the methanol column density of $2 \times 10^{19}$ cm$^{-2}$ provided by \citet{Motiyenko20} (see their Table 3).

Around 30\% of acetaldehyde is formed on the grains prior to the desorption of the ices, of which around one-third is formed at temperatures below 20~K, mainly through surface chemistry as the ices build up. 
Reactions between CH$_2$ and CO are important in producing ketene, CH$_2$CO, which is further hydrogenated to acetaldehyde on the grain surfaces by mobile atomic H. 
We would also expect this cold, grain-surface mechanism to be relevant to  acetaldehyde production on grains in G+0.693, where the dust temperature is $<$30 K \citep{rodriguez-fernandez2004}.

In the hot-core models, some CH$_3$CHO is also formed at intermediate temperatures ($\sim$20--100~K), as the result of photo-processing of the ices by cosmic ray-induced UV photons. 
However, approximately two thirds of total acetaldehyde production during the entire hot core evolution is calculated to occur in the gas phase, through the barrier-less reaction O + C$_2$H$_5$ $\rightarrow$ CH$_3$CHO + H 
\citep{Tsang86}.
This reaction is fed by the release of trapped solid-phase C$_2$H$_5$ into the gas when the ices sublimate at temperatures greater than 100~K, while atomic O is a product of the gas-phase destruction of more stable species such as CO$_2$ that are released at around the same time. The dominance of the gas-phase reaction in producing acetaldehyde results in a delay between the sublimation of the dust-grain ice mantles and the attainment of the peak CH$_3$CHO abundance. Furthermore, due to the gradually increasing temperature in the model, this delay in peak abundance occurs at an elevated temperature of 257~K, which is in good agreement with the rotational temperature used in the spectral model of Sgr B2(N1S).

The chemical model indicates that propanal should reach a peak gas-phase abundance around 60 times lower than that of acetaldehyde (although the {fast} warm-up timescale model suggests a value closer to 10). 
This is consistent with the observational upper limits derived toward Sgr~B2(N1S) using a rotational temperature of 250~K (we note that the chemical models do not explicitly distinguish conformers). The excitation temperature used to determine the propanal upper limit is chosen to be consistent with acetaldehyde.

Unlike the treatment for acetaldehyde, the chemical model includes only grain-surface mechanisms for the production of propanal; this solid-phase material is preserved on the grains at low temperatures, and then released into the gas phase when dust temperatures are sufficiently high to allow the thermal desorption of propanal and of other grain-surface species.
Due to its direct release into the gas phase, the abundance of propanal reaches its peak at an ambient gas temperature of $\sim$170~K, substantially lower than the 250~K assumed here for its excitation temperature, and lower than the 257~K at which acetaldehyde reaches its peak in the model. In order to test the influence on observational results of this lower temperature, which is predicated on a uniquely grain-surface origin for propanal, an excitation temperature of 170~K was also tested in the spectral analysis of Sgr~B2(N1S) (see bottom part of Table \ref{t:coldens}). The model results nevertheless remain consistent with observations.

We note, however, that a a gas-phase formation route for propanal may in fact be plausible; 
\citet{Tsang88}
recommends a rate of $1.6 \times 10^{-10}$ cm$^{3}$s$^{-1}$ for the O + \textit{n}-C$_3$H$_7$ reaction, identical to the recommended rate of the equivalent process for acetaldehyde production. Formation of an aldehyde is the preferred outcome in either reaction. The same rate was also proposed by 
\citet{Tsang88} 
for the reaction of the related radical \textit{i}-C$_3$H$_7$, leading to acetone production.

With such a reaction for propanal included in the astrochemical model, some enhancement in the peak gas-phase abundance would be expected.
By comparing the behavior of acetaldehyde and the other species in the model, it is possible to make a crude prediction as to the strength of influence of this mechanism. The models predict \textit{n}-C$_3$H$_7$ in the ice to be around five times less abundant than C$_2$H$_5$ immediately prior to ice sublimation, suggesting a five times lower base rate of gas-phase propanal production versus acetaldehyde. To gauge the importance of this lower rate, it is also noteworthy that only some of the gas-phase acetaldehyde is formed through the gas-phase reaction, with the rest coming directly from the grain surfaces. On this basis it might be reasonable to expect that if gas-phase production of propanal were to dominate over any grain-surface mechanism, it would reach a maximum abundance of $\sim$10\% that of acetaldehyde. This crude estimate also is approximately in keeping with the upper limits toward Sgr~B2(N1S).

The question then arises as to whether such a gas-phase mechanism could also produce butanal. On the assumption that it could -- through the reaction of \textit{n}-C$_4$H$_9$ with atomic oxygen -- then the abundance of butanal could be comparable to that of propanal. All four of the C$_4$H$_9$ radicals in the network (i.e., the straight-chain and branched forms, with the radical site on either the primary or secondary carbon) achieve ice abundances similar to the C$_3$H$_7$ radicals. Two of these might be expected to form ketones rather than aldehydes. The other two, which would produce either n- or i-butanal, reach very similar abundances, indicating no obvious distinction between the two forms.

While gas-phase production of aldehydes may be plausible in hot cores such as Sgr B2(N1S), for G+0.693 -- which is assumed to have been heated by a shock, releasing the ice mantles -- the timescales available may be too short to have enabled substantial gas-phase chemistry to occur. In the {fast} warm-up model of 
\cite{Garrod21}, 
which has the shortest desorption timescale of their three hot-core models and the weakest gas-phase contribution to acetaldehyde production, a period on the order of 10$^4$ yr is required for gas-phase CH$_3$CHO production to become significant following ice sublimation. The gas density is also set to a relatively high value of $n_\mathrm{H}=2 \times 10^8$ cm$^{-3}$, meaning that gas-phase chemistry occurs quite rapidly when the ice mantles are released. 

If the passage of a shock through G+0.693 led to compression of the gas to a density lower than the above value, then the gas-phase acetaldehyde production timescale would be longer than in the hot core model (assuming that the gas-phase production mechanisms were the same). In fact, the gas density in G+0.693 is lower, in the range 10$^{4}$-10$^{5}$ cm$^{-3}$ \citep{zeng2020}, which certainly does not favor gas-phase chemistry on short timescales.
The presence of gas-phase COMs might instead be more directly related to the shock-induced sputtering of the ice mantles, releasing aldehydes into the gas phase that were already formed in the solid phase. The shock velocities are expected to be between 15-20 km s$^{-1}$, which would be sufficient to fully erode the icy mantles of dust grains via sputtering (see \citealt{jimenez2008}). 


If the acetaldehyde detected toward G+0.693 was formed entirely on the grains, followed by a rapid shock-induced release, then the cold, grain-surface abundances derived from the early stages of the hot-core models may be comparable to the observed gas-phase abundances. The models suggest a purely solid-phase ratio of acetaldehyde to propanal of $\sim$7.7. This is in excellent agreement with the observational ratio of $\sim$7. However, it is unclear how the shock processing might further affect the subsequent gas-phase molecular abundances, so this solid-phase ratio might not be accurately reflected in the gas-phase abundances of the two aldehydes.

Regarding the possible butanal abundance in G+0.693, one may crudely consider the ratio of C$_2$H$_5$ radicals to C$_3$H$_7$ radicals within the bulk ices during the cold stages of the model (i.e., the cold collapse stage before the warm-up begins) as a proxy for the propanal to butanal ratio within the ice. In this view, production of C$_2$H$_5$CHO and C$_3$H$_7$CHO would occur through grain-surface or bulk-ice radical-addition reactions between HCO and either C$_2$H$_5$ or C$_3$H$_7$. The models suggest a C$_2$H$_5$:C$_3$H$_7$ ratio around 8 in the ice, which again comes very close to the observed upper limit for C$_2$H$_5$CHO:C$_3$H$_7$CHO.

Thus, based on estimates derived from the astrochemical models, the observational upper limits on both propanal and butanal could be very close to their true abundances toward Sgr B2(N1S) and toward G+0.693. In Sgr B2(N1S), this assumes that gas-phase production is strong. Weak gas-phase production could produce values rather lower than the upper limits. In G+0.693, the observed ratios are well reproduced without the need for gas-phase chemistry. 
However, we note that these ratios do not directly indicate the fractional abundances -- the quantities of aldehydes produced relative to total hydrogen may vary between the two sources. Indeed the column densities measured toward Sgr B2(N1S) and G+0.693 are very different; G+0.693 may have overall lower fractional abundances of all aldehydes while maintaining a higher ratio of the larger homologs, as compared with Sgr B2(N1S).





The inclusion of these molecules in a more comprehensive gas-grain chemical network would be highly valuable in providing more accurate predictions for future observational searches.
\\



\section{Conclusions}
\label{s:conclusions}
We are finally starting to fill the gap between observational and laboratory data that has existed for several decades. In this context, we have investigated and analyzed the millimeter- and submillimeter-wave spectrum of \textit{n}- and \textit{i}-butanal from 75 to 325 GHz. With our experimental data set we can predict the rotational transitions for both isomers with narrow line widths (full width at half maximum, FWHM) of less than 0.1 km~s${^{-1}}$ at 100 GHz, expressed in its equivalent radial velocity, which is more than enough to perform a confident astronomical search for these species. We employed the newly determined rotational parameters to prepare accurate line catalogs, which we used to search for the molecule toward the hot core Sgr B2(N1S) and the G+0.693-0.027 molecular cloud. We complemented these searches toward both sources with a comparison to shorter aldehydes. We obtained the following results:
\begin{enumerate}
\item We report a non-detection of \textit{n}- and \textit{i}-butanal toward Sgr B2(N1S) 
with ALMA. Depending on the assumed temperature, we find that propanal, \textit{n}-butanal, and \textit{i}-butanal are at least 5--11, 2--6, and 6-18 times less abundant 
than acetaldehyde, respectively.
\item We also report a non-detection of both isomers of butanal toward G+0.693 based on Yebes 
40m and IRAM 30m observations. We find that propanal is seven times less 
abundant than acetaldehyde, while the propanal / \textit{ct-n}-butanal ratio is 
> 9 and the acetaldehyde / \textit{ct-n}-butanal ratio is > 63. 

\item Finally, our astrochemical models suggest the observational upper limits on both 
propanal and butanal could be close to their true abundances
toward Sgr B2(N1S) and G+0.693. This assumes that gas-phase production of aldehydes is dominant in Sgr B2(N1S), 
while in G+0.693 aldehyde production on dust grains is sufficient to reproduce observed molecular ratios. If gas-phase mechanisms are not effective in Sgr B2(N1S), then the true values could be substantially lower than the upper limits observed here.
\end{enumerate}

The most stringent constraints on the relative abundances of the three aldehydes in the ISM were obtained for G+0.693-0.027. Our results show that with each incremental increase in the size of the alkyl group in the aldehyde molecule, there is a drop of approximately one order of magnitude in abundance.


\begin{acknowledgements}

The authors thank the financial fundings from Ministerio de Ciencia e Innovaci{\'o}n (CTQ2016- 76393-P, PID2019-111396GB-I00 and PID2020-117742GB-I00), the Ministerio de Economia Industria y Competitividad (Grant AYA2017-87515-P)  Junta de 
Castilla y Le{\'o}n (Grants VA010G18 and Grant VA244P20), and the European Research Council under the European Union’s Seventh Framework Programme (FP/2007–2013)/ERC-2013-SyG, Grant Agreement no. 610256 NANOCOSMOS. M.S.N. 
acknowledges funding from the Spanish “Ministerio de Ciencia, Innovaci{\'o}n y Universidades” under predoctoral FPU Grant 
(FPU17/02987).
V.M.R. has received funding from the Comunidad de Madrid through the Atracci\'on de Talento Investigador (Doctores con experiencia) Grant (COOL: Cosmic Origins Of Life; 2019-T1/TIC-15379). I.J.-S. and J.M.-P. have received partial support from the Spanish State Research Agency (AEI) through project numbers PID2019-105552RB-C41 and MDM-2017-0737 Unidad de Excelencia "Maria de Maeztu”- Centro de Astrobiolog\'a (CSIC-INTA).

This paper makes use of the following ALMA data: 
ADS/JAO.ALMA\#2016.1.00074.S. 
ALMA is a partnership of ESO (representing its member states), NSF (USA), and 
NINS (Japan), together with NRC (Canada), NSC and ASIAA (Taiwan), and KASI 
(Republic of Korea), in cooperation with the Republic of Chile. The Joint ALMA 
Observatory is operated by ESO, AUI/NRAO, and NAOJ. The interferometric data 
are available in the ALMA archive at https://almascience.eso.org/aq/.
Part of this work has been carried out within the Collaborative
Research Centre 956, sub-project B3, funded by the Deutsche
Forschungsgemeinschaft (DFG) -- project ID 184018867.
RTG acknowledges support from the National Science Foundation (grant No. AST 19-06489).

\end{acknowledgements}

\begin{appendix}
\label{appendix}

\section{Complementary tables}
In Table ~\ref{TableA1} we provide a list with the theoretical frequencies of the lowest vibrational modes of \textit{n}- and \textit{i}-PrCHO.

\begin{table*}[!ht]
\begin{center}
\caption{Theoretically predicted harmonic vibrational frequencies\tablefootmark{(a)} of the vibrational modes for the lowest-energy conformers of \textit{n}- and \textit{i}-PrCHO.}
\label{TableA1}
\vspace*{0.0ex}
\begin{tabular}{clllllll}
\hline\hline
\multicolumn{1}{c}{vibrational} & \multicolumn{4}{c}{\textit{n}-PrCHO} & \multicolumn{2}{c}{\textit{i}-PrCHO} \\ 
\multicolumn{1}{c}{mode} & \multicolumn{1}{c}{\textit{cis-gauche}} & \multicolumn{1}{c}{\textit{cis-trans}} & \multicolumn{1}{c}{\textit{gauche-gauche}} & \multicolumn{1}{c}{\textit{gauche-trans}} & \multicolumn{1}{c}{\textit{gauche}} & \multicolumn{1}{c}{\textit{trans}} \\ 
\hline
1 & 109.5 & 82.5 & 66.3 & 67.2 & 95.3 & 63.4 \\
2 & 162.8 & 172.6 & 116.6 & 98.3 & 204.9 & 208.4 \\ 
3 & 201.0 & 196.2 & 222.3 & 236.1 & 238.6 & 236.2 \\ 
4 & 283.6 & 248.2 & 304.4 & 255.5 & 275.4 & 331.0 \\
5 & 364.9 & 347.9 & 402.2 & 385.7 & 341.6 & 331.3 \\ 
6 & 664.5 & 682.0 & 515.5 & 518.9 & 399.9 & 352.4 \\ 
7 & 721.2 & 703.8 & 771.2 & 753.4 & 649.9 & 551.8 \\ 
8 & 802.2 & 799.0 & 795.6 & 821.9 & 816.5 & 858.4 \\ 
9 & 863.3 & 873.7 & 917.4 & 917.5 & 926.9 & 938.4 \\ 
10 & 959.9 & 960.1 & 961.8 & 964.7 & 946.8 & 946.9 \\
11 & 980.2 & 979.0 & 993.3 & 1030.2 & 960.0 & 967.8 \\
12 & 1073.5 & 1074.9 & 1087.3 & 1078.6 & 991.4 & 1006.7 \\ 
13 & 1138.0 & 1155.4 & 1143.6 & 1158.0 & 1149.0 & 1163.5 \\ 
14 & 1166.7 & 1174.0 & 1180.1 & 1182.6 & 1181.1 & 1203.3 \\ 
15 & 1270.7 & 1272.4 & 1256.2 & 1270.7 & 1221.9 & 1206.3 \\ 
16 & 1316.2 & 1331.7 & 1299.7 & 1305.3 & 1335.5 & 1339.3 \\ 
17 & 1390.2 & 1341.2 & 1371.3 & 1344.2 & 1384.8 & 1365.3 \\ 
18 & 1412.4 & 1424.1 & 1394.0 & 1406.3 & 1422.6 & 1417.8 \\
19 & 1432.6 & 1436.5 & 1438.6 & 1434.6 & 1429.9 & 1431.9 \\ 
20 & 1443.6 & 1442.3 & 1445.8 & 1447.3 & 1451.7 & 1444.9 \\ 
21 & 1471.7 & 1476.6 & 1486.0 & 1489.4 & 1505.2 & 1507.3 \\ 
22 & 1505.8 & 1510.7 & 1511.1 & 1513.4 & 1510.6 & 1512.5 \\ 
23 & 1519.7 & 1521.4 & 1521.8 & 1520.2 & 1524.2 & 1524.7 \\ 
24 & 1530.8 & 1527.9 & 1524.5 & 1527.9 & 1528.9 & 1530.6 \\ 
25 & 1836.5 & 1836.7 & 1839.4 & 1839.0 & 1836.4 & 1840.1 \\
26 & 2947.7 & 2947.3 & 2926.8 & 2934.1 & 2935.6 & 2916.4 \\
27 & 3041.0 & 3037.6 & 3051.5 & 3046.4 & 3034.0 & 3050.7 \\
28 & 3054.1 & 3049.4 & 3054.2 & 3053.2 & 3052.5 & 3053.0 \\
29 & 3067.0 & 3065.9 & 3058.0 & 3053.6 & 3064.3 & 3098.7 \\
30 & 3076.8 & 3076.6 & 3096.9 & 3086.1 & 3123.0 & 3121.2 \\
31 & 3107.5 & 3102.8 & 3123.1 & 3121.8 & 3133.8 & 3128.2 \\
32 & 3125.3 & 3126.3 & 3128.1 & 3130.1 & 3137.3 & 3130.8 \\
33 & 3145.8 & 3129.0 & 3132.5 & 3131.4 & 3147.3 & 3133.7 \\
\hline 
\end{tabular}
\end{center}
\vspace*{-2.5ex}
\tablefoot{\tablefootmark{(a)}Frequencies (in cm$^{-1}$) calculated at the CCSD/cc-pVTZ level of theory. They are needed to compute the vibrational contribution to the partition function.}
\end{table*}

\clearpage

\section{Complementary figures: Spectra and population diagram of Sgr~B2(N1S)}
\label{a:spectra}

\begin{figure*}
\centerline{\resizebox{0.82\hsize}{!}{\includegraphics[angle=0]{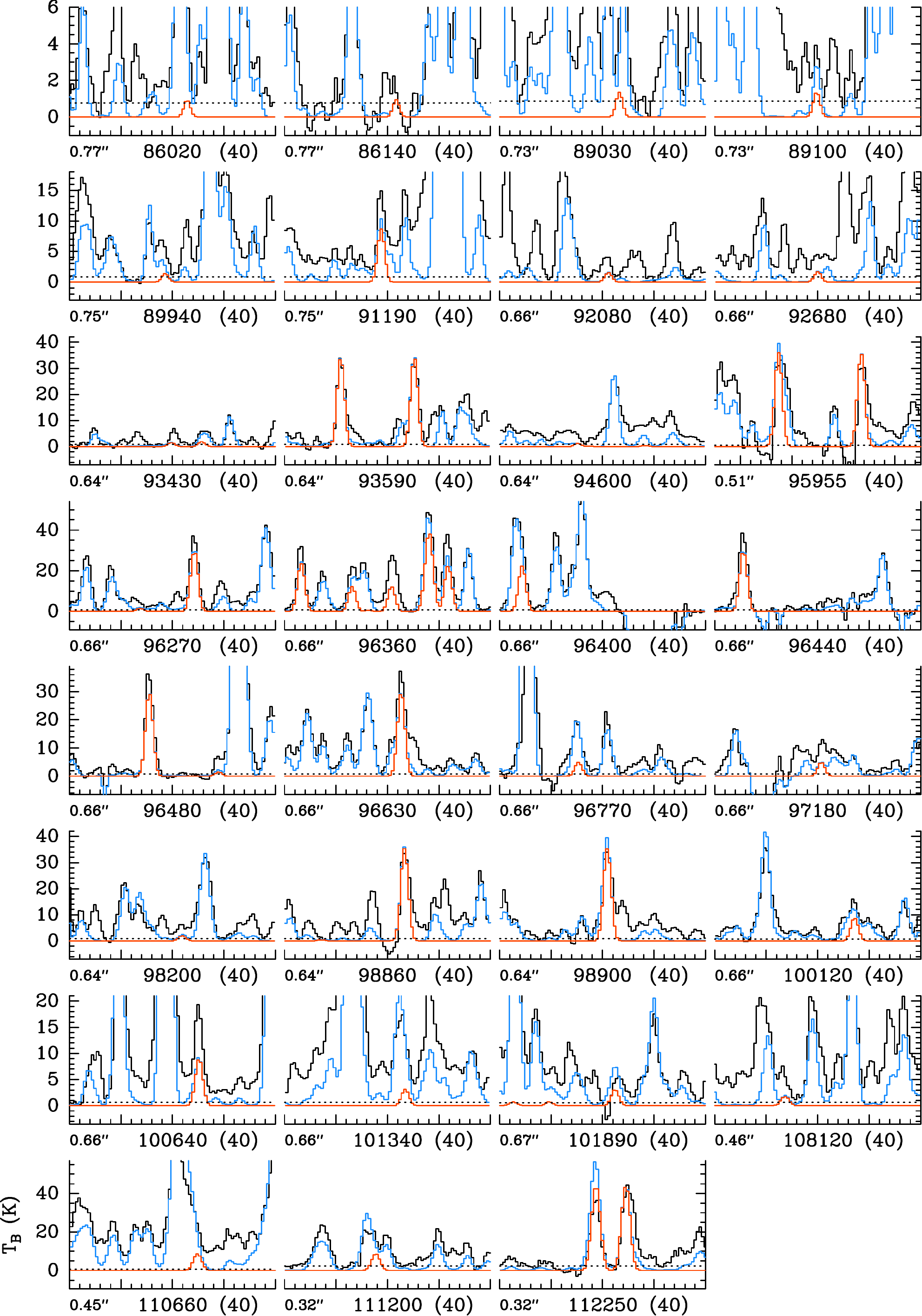}}}
\caption{Transitions of CH$_3$CHO, $\varv = 0$ covered by the ReMoCA  
survey. The best-fit LTE synthetic spectrum of CH$_3$CHO, $\varv = 0$ 
is displayed in red and overlaid on the observed spectrum of Sgr~B2(N1S), which is shown 
in black. The blue synthetic spectrum contains the contributions of all 
molecules identified in our survey so far, including the species shown in red. 
The central frequency and width are indicated in MHz below each panel, as is the half-power beam width. The 
y axis is labeled in brightness temperature units (K). The dotted line 
indicates the $3\sigma$ noise level.
}
\label{f:spec_ch3cho_ve0}
\end{figure*}

\begin{figure*}
\centerline{\resizebox{0.82\hsize}{!}{\includegraphics[angle=0]{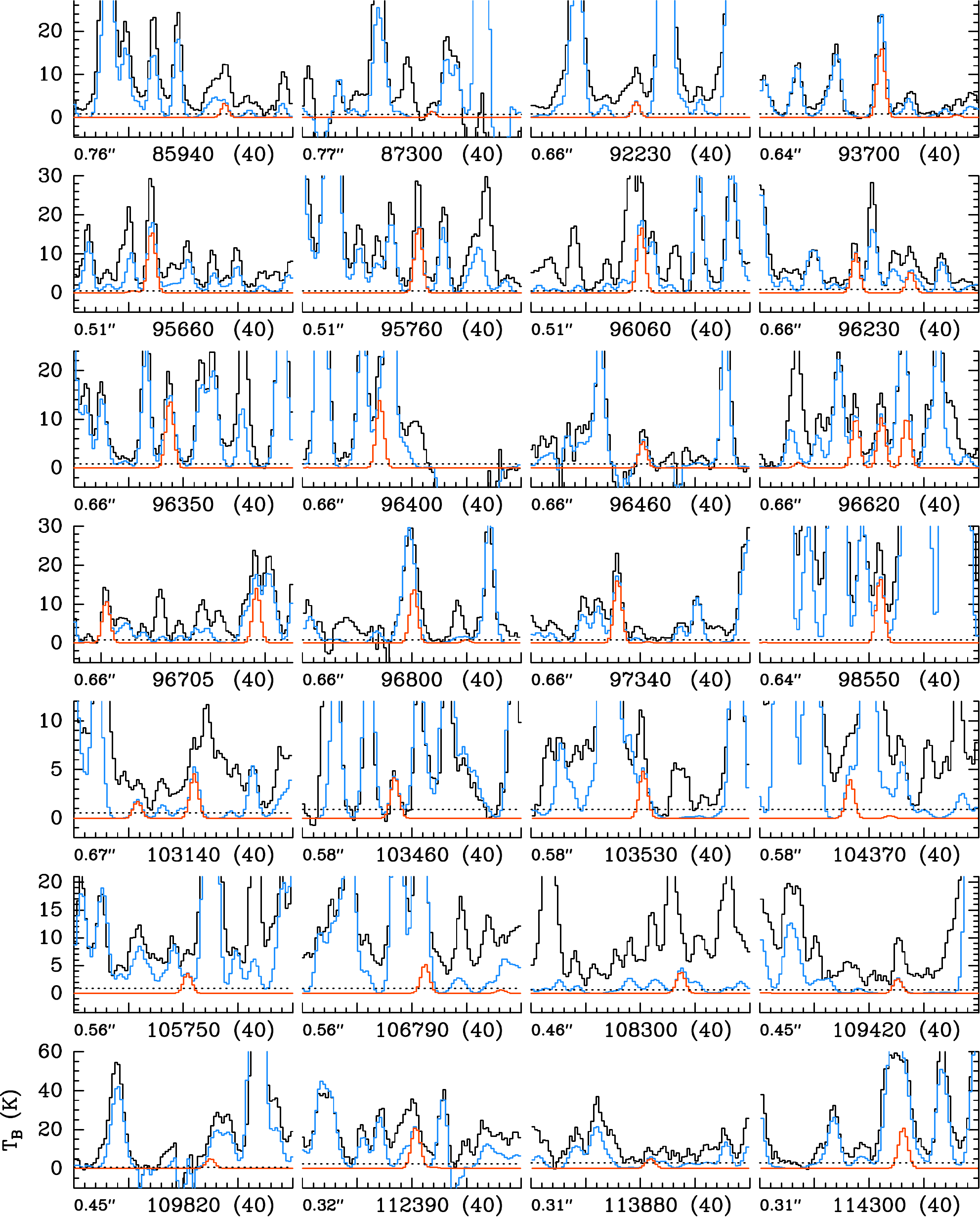}}}
\caption{Same as Fig.~\ref{f:spec_ch3cho_ve0} but for CH$_3$CHO, $\varv$=1.}
\label{f:spec_ch3cho_ve1}
\end{figure*}

\begin{figure*}
\centerline{\resizebox{0.82\hsize}{!}{\includegraphics[angle=0]{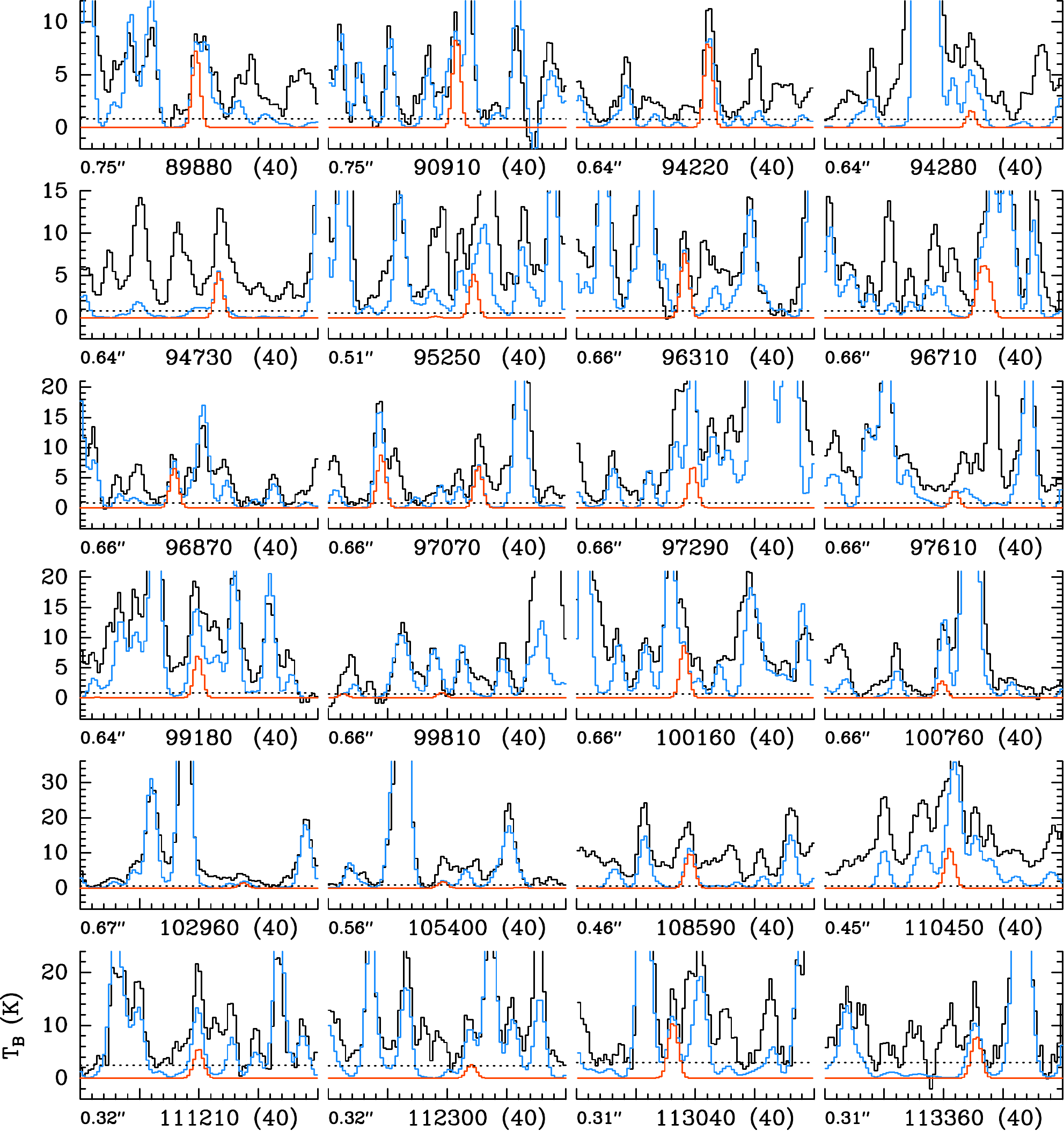}}}
\caption{Same as Fig.~\ref{f:spec_ch3cho_ve0} but for CH$_3$CHO, $\varv$=2.}
\label{f:spec_ch3cho_ve2}
\end{figure*}

\begin{figure}[!ht]
\centerline{\resizebox{0.95\hsize}{!}{\includegraphics[angle=0]{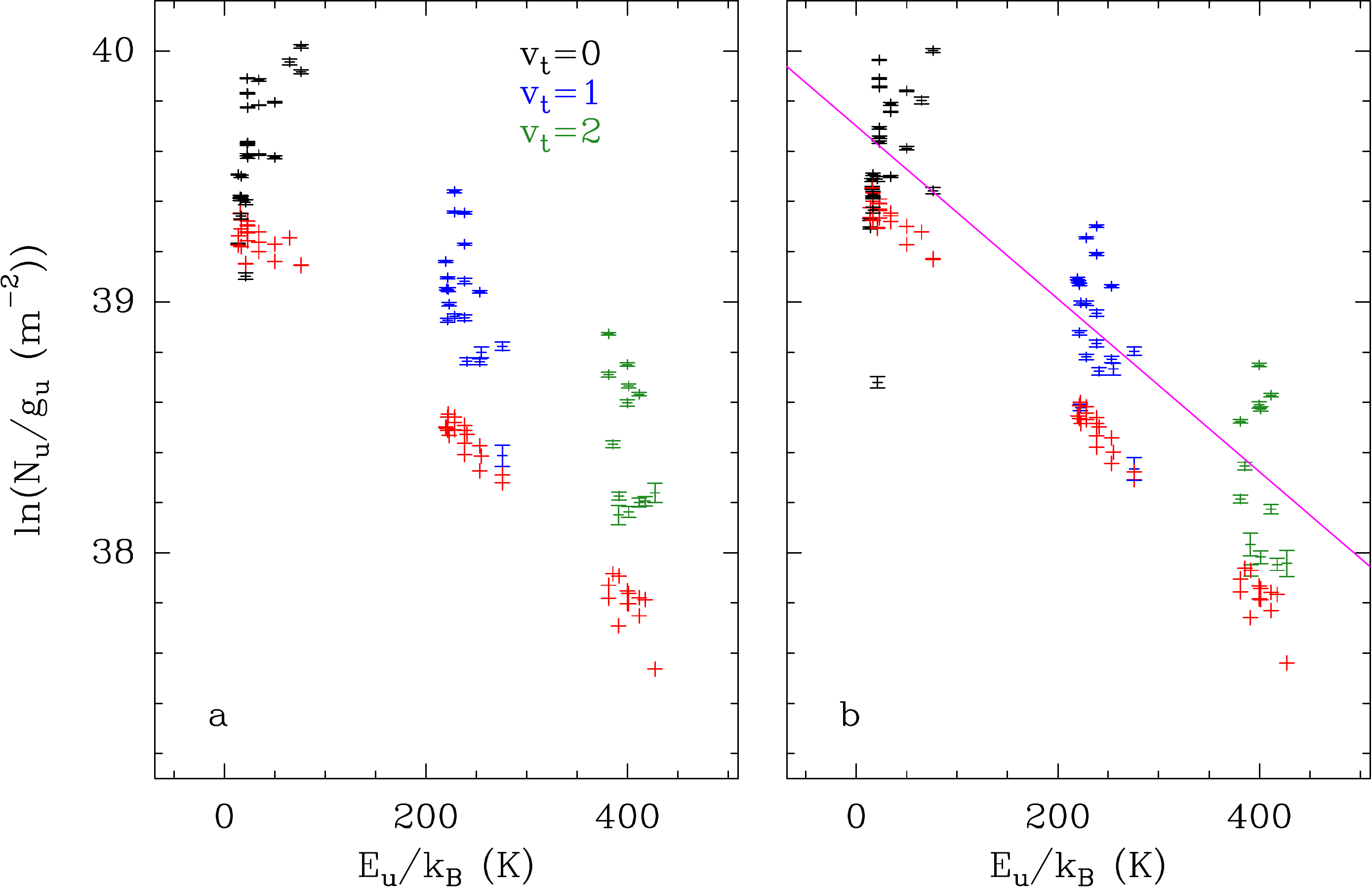}}}
\caption{Population diagram of CH$_3$CHO toward Sgr~B2(N1S). The 
observed data points are shown in various colors (but not red), as indicated in 
the upper-right corner of panel \textbf{a}, and the synthetic populations are 
shown in red. No correction is applied in panel \textbf{a}. 
In panel \textbf{b}, the optical depth correction has been applied to both the 
observed and synthetic populations, and the contamination by all other 
species included in the full model has been removed from the observed 
data points. The purple line is a linear fit to the observed populations (in 
linear-logarithmic space).
}
\label{f:popdiag_ch3cho}
\end{figure}

\begin{figure*}
\centerline{\resizebox{0.82\hsize}{!}{\includegraphics[angle=0]{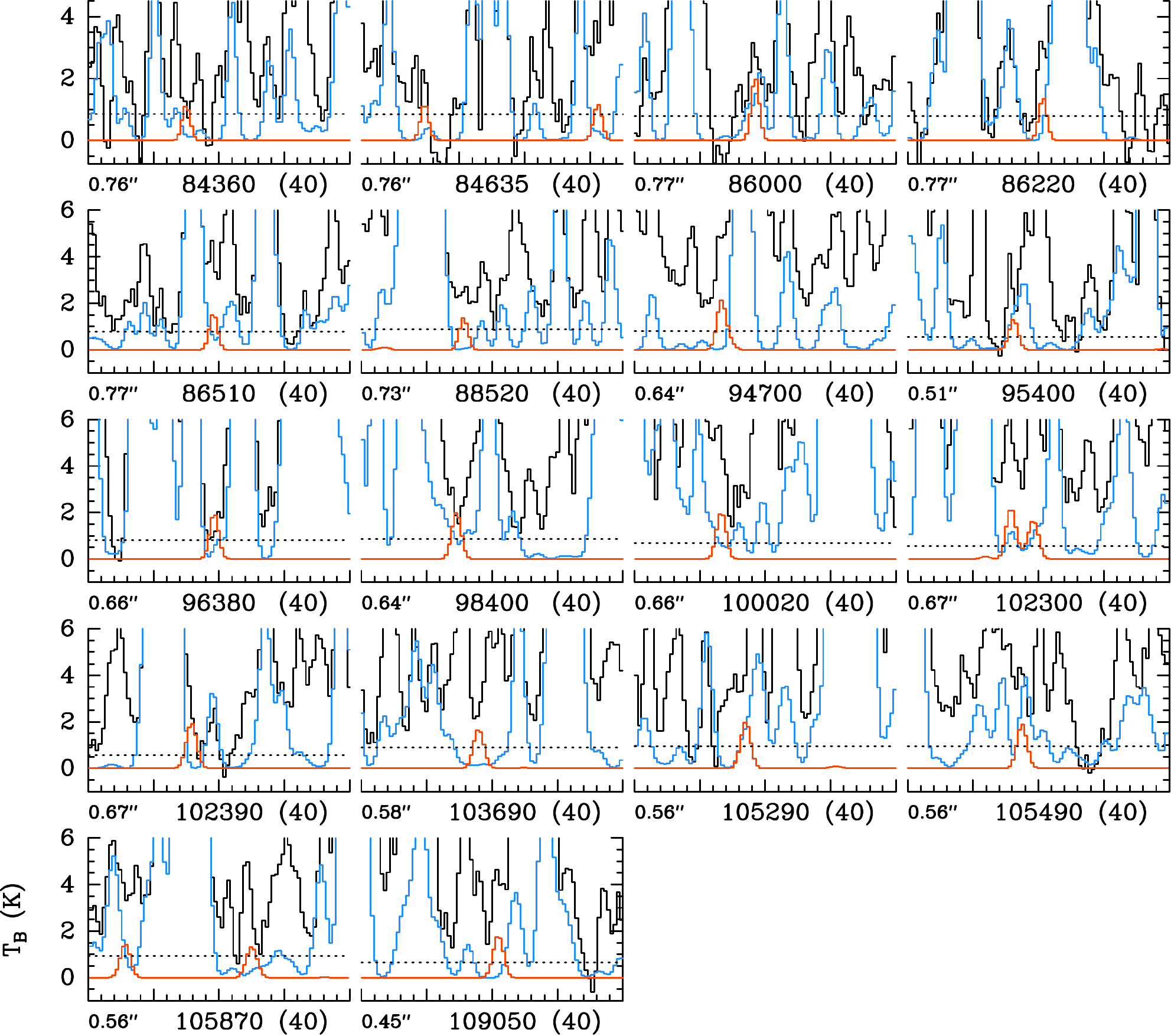}}}
\caption{Same as Fig.~\ref{f:spec_ch3cho_ve0} but here the red spectrum shows the
synthetic spectrum of \textit{syn}-C$_2$H$_5$CHO, $\varv$=0 used to derive the 
upper limit on its column density reported in Table~\ref{t:coldens}. The 
blue spectrum does not contain the contribution of the species shown in red.}
\label{f:spec_c2h5cho-s_ve0}
\end{figure*}

\begin{figure*}
\centerline{\resizebox{0.82\hsize}{!}{\includegraphics[angle=0]{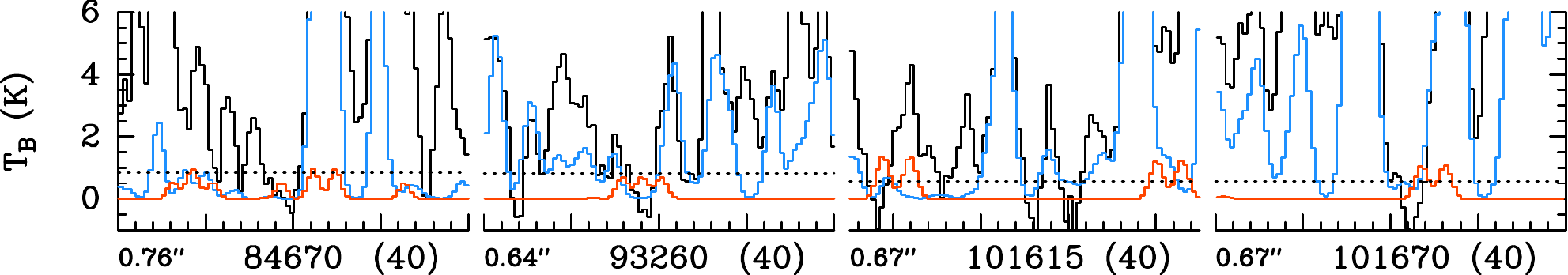}}}
\caption{Same as Fig.~\ref{f:spec_c2h5cho-s_ve0} but for 
\textit{gauche-}C$_2$H$_5$CHO, $\varv$=0.}
\label{f:spec_c2h5cho-g_ve0}
\end{figure*}

\begin{figure*}
\centerline{\resizebox{0.82\hsize}{!}{\includegraphics[angle=0]{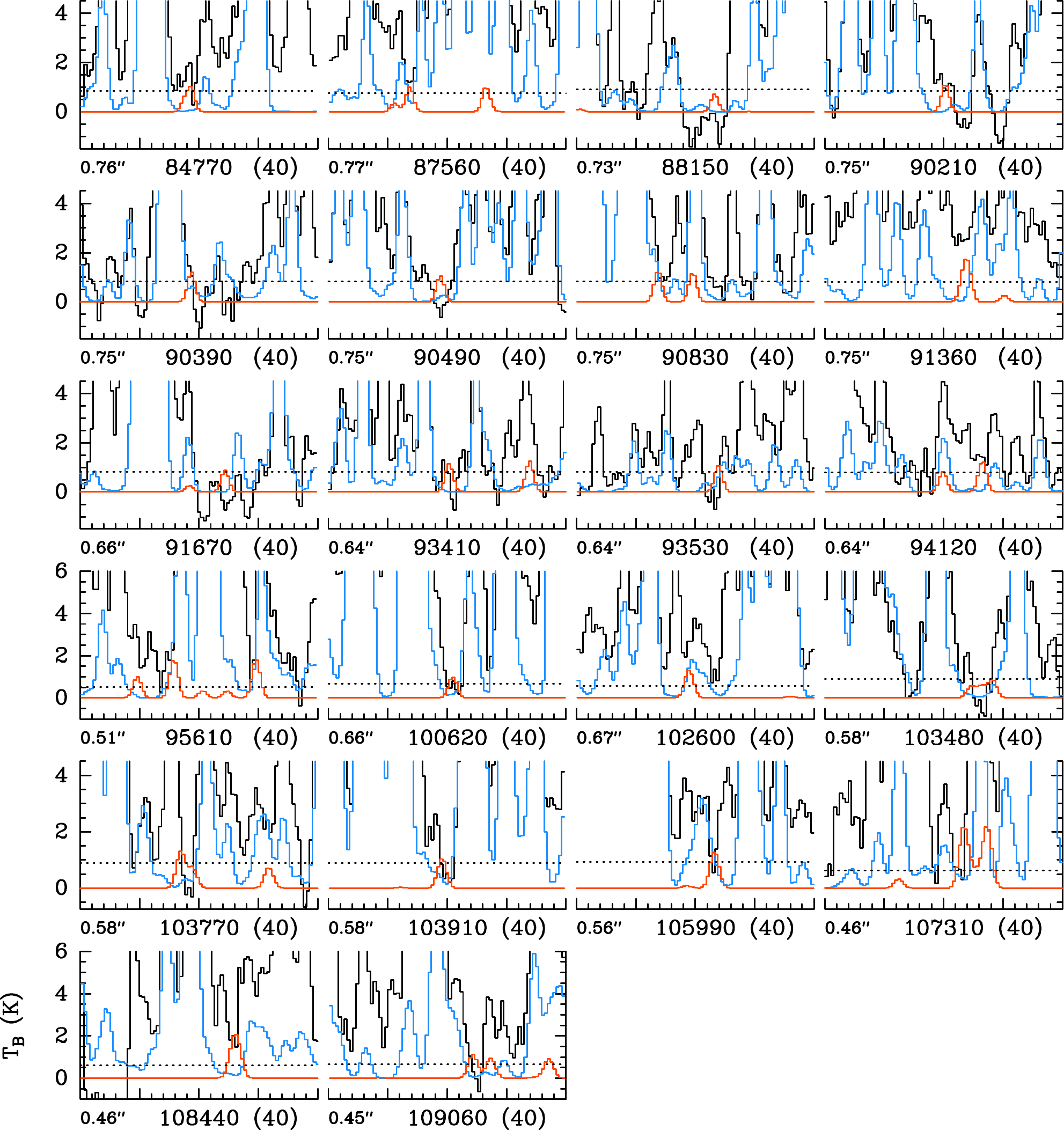}}}
\caption{Same as Fig.~\ref{f:spec_c2h5cho-s_ve0} but for 
\textit{cis-gauche} \textit{normal-}C$_3$H$_7$CHO, $\varv$=0.}
\label{f:spec_c3h7cho-cg_ve0}
\end{figure*}

\begin{figure*}
\centerline{\resizebox{0.82\hsize}{!}{\includegraphics[angle=0]{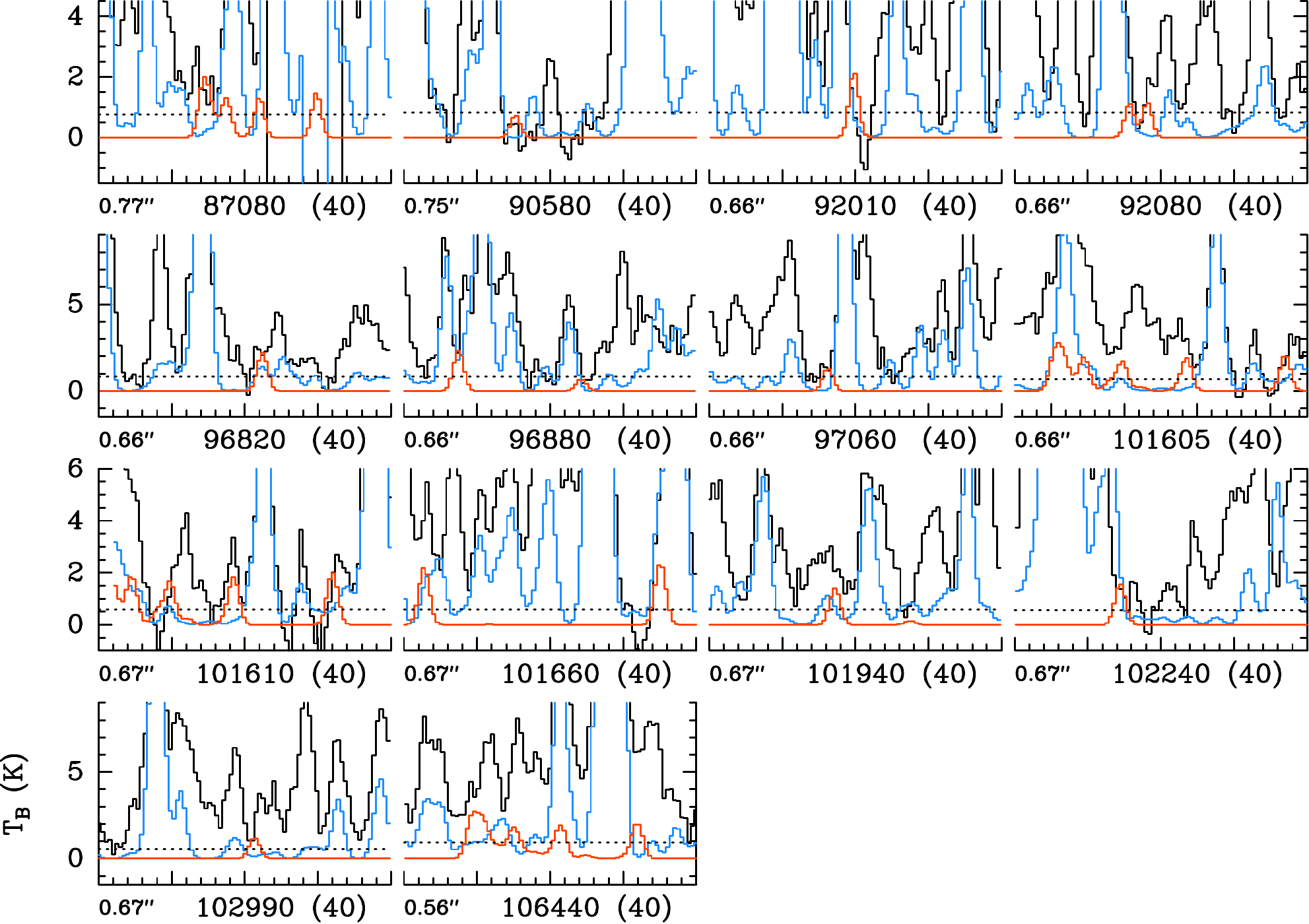}}}
\caption{Same as Fig.~\ref{f:spec_c2h5cho-s_ve0} but for 
\textit{cis-trans} \textit{normal-}C$_3$H$_7$CHO, $\varv$=0.}
\label{f:spec_c3h7cho-ct_ve0}
\end{figure*}

\begin{figure*}
\centerline{\resizebox{0.82\hsize}{!}{\includegraphics[angle=0]{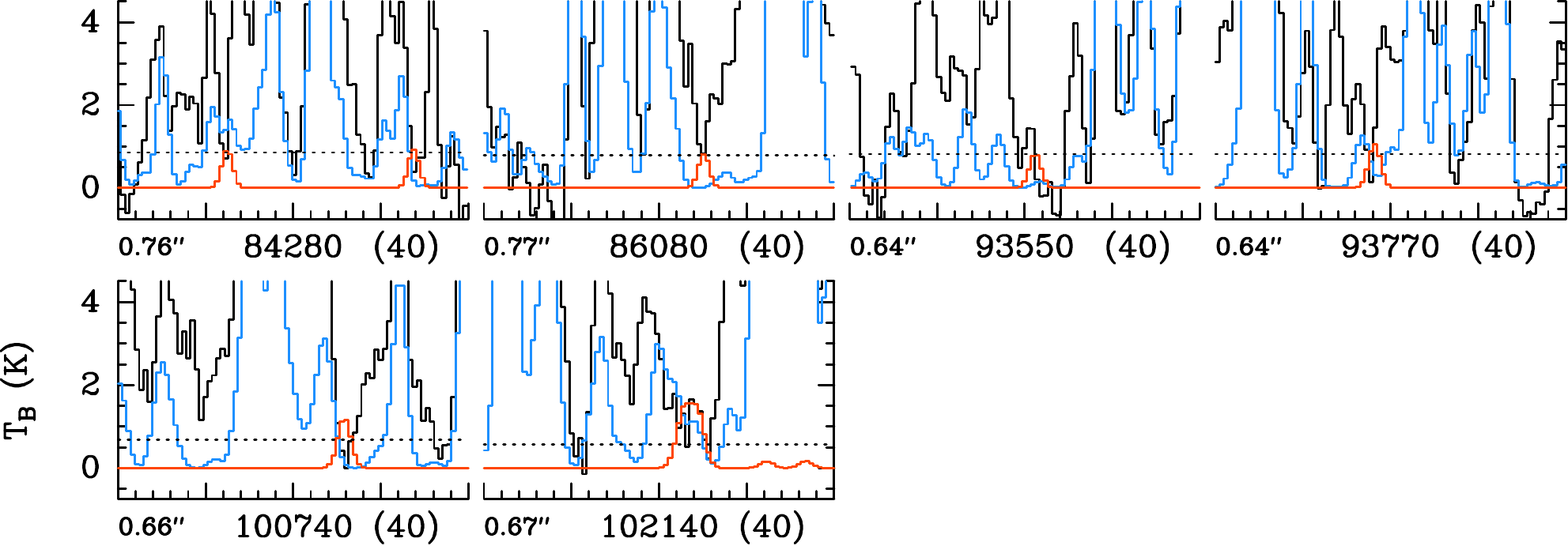}}}
\caption{Same as Fig.~\ref{f:spec_ch3-2-chcho-g_ve0} but for 
\textit{gauche} \textit{iso-}C$_3$H$_7$CHO, $\varv$=0.}
\label{f:spec_ch3-2-chcho-g_ve0}
\end{figure*}

Figures~\ref{f:spec_ch3cho_ve0}--\ref{f:spec_ch3cho_ve2} show the
transitions of CH$_3$CHO, $\varv_{\rm t}$=0, $\varv_{\rm t}=1$, and 
$\varv_{\rm t}=2$ that are covered 
by the ReMoCA  survey and contribute significantly to the signal detected 
toward Sgr~B2(N1S). Figure~\ref{f:popdiag_ch3cho} shows the population diagram
of CH$_3$CHO for Sgr~B2(N1S). Figures~\ref{f:spec_c2h5cho-s_ve0} and 
\ref{f:spec_c2h5cho-g_ve0} illustrate the non-detection of 
\textit{syn} and \textit{gauche} C$_2$H$_5$CHO, respectively,
Figs~\ref{f:spec_c3h7cho-cg_ve0} and \ref{f:spec_c3h7cho-ct_ve0} the 
non-detection of the \textit{cis-gauche} and \textit{cis-trans} conformers of 
\textit{normal}-C$_3$H$_7$CHO, respectively, and 
Fig.~\ref{f:spec_ch3-2-chcho-g_ve0} the non-detection of
\textit{gauche} \textit{iso-}C$_3$H$_7$CHO, all toward Sgr~B2(N1S). 

\section{Analysis of acetaldehyde (CH$_3$CHO) toward the G+0.693$-$0.027 molecular cloud}
\label{app-ch3cho}

\begin{figure*}
\centerline{\resizebox{1.0\hsize}{!}{\includegraphics[angle=0]{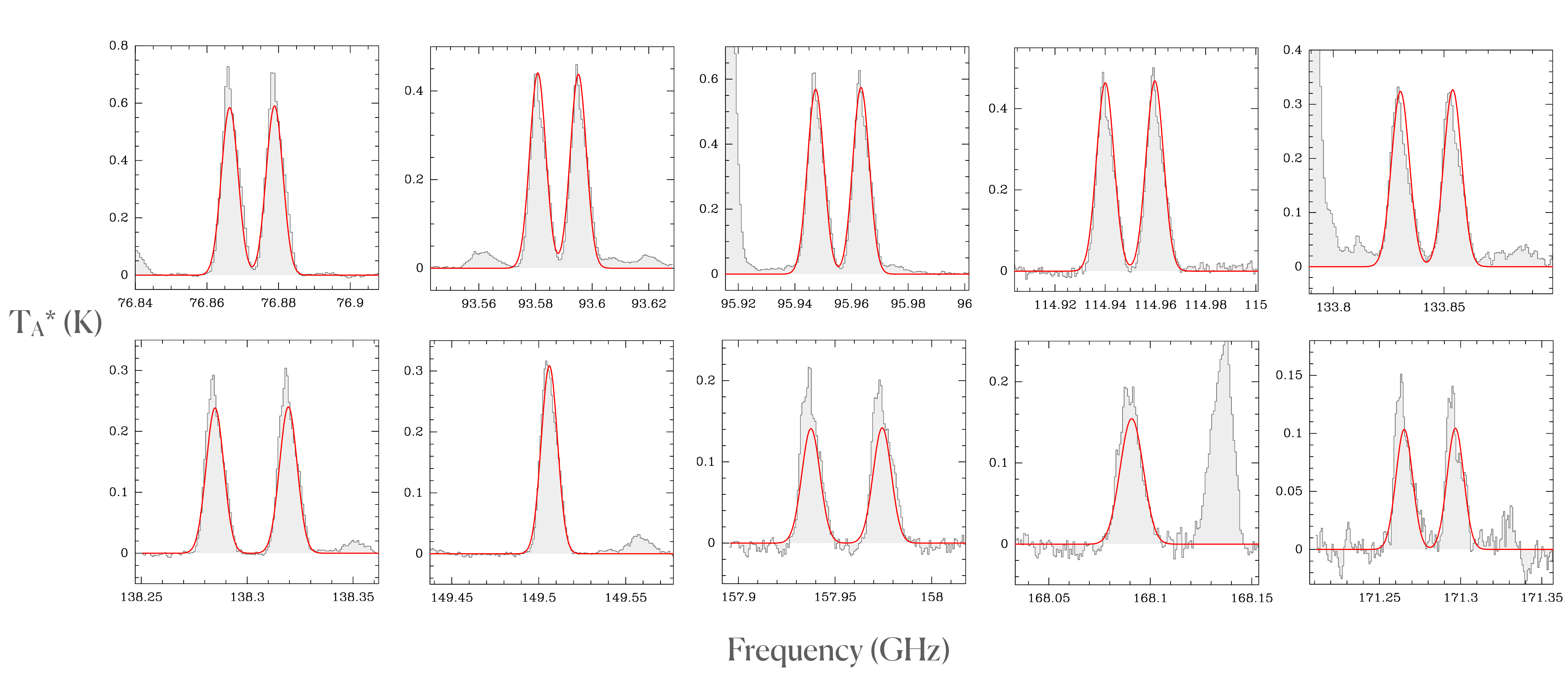}}}
\caption{Selected unblended transitions of CH$_3$CHO (see Table \ref{tab:ch3cho}) detected toward G+0.693$-$0.027 used to perform the LTE analysis. The gray histograms show the observed spectra, and the red curve is the best LTE fit obtained with MADCUBA.}
\label{f:ch3cho_go693}
\end{figure*}

To perform the LTE analysis of CH$_3$CHO, we selected transitions that are not blended with other molecular species and that cover a broad range of energy levels in order to constrain the excitation conditions. The transitions used in the analysis are listed in Table \ref{tab:ch3cho}. We used the spectroscopic predictions of \citet{Smirnov14}, as in the analysis of Sgr B2(N1S). We used MADCUBA$-$SLIM, leaving as free parameters the molecular column density ($N$), the rotational temperature $T_{\rm rot}$, the velocity (V$_{\rm LSR}$) and linewidth $\Delta$V. The derived physical parameters are shown in Table \ref{tab:g0693}, and the fits of the transitions are shown in Fig. \ref{f:ch3cho_go693}.

\begin{table}
\centering
\tabcolsep 6pt
\caption{Selected unblended transitions of CH$_3$CHO detected toward G+0.693$-$0.027 used to perform the LTE analysis.}
\vspace{2mm}
\begin{tabular}{ c c c  c c }
\hline
 Frequency & Transition    & log  $I$   & $g_{\rm u}$ & E$_{\rm up}$ \\
 (MHz) &   &  (nm$^2$ MHz) & & (K)    \\
\hline
76866.4396   & 4$_{0,4}-$3$_{0,3}$ E  & -4.7276  & 9   & 9.33    \\ 
76878.9554   & 4$_{0,4}-$3$_{0,3}$ A  & -4.7273  & 9   & 9.23    \\ 
93580.9130   & 5$_{1,5}-$4$_{1,4}$ A  & -4.4871  & 11  & 15.75    \\ 
93595.2373   & 5$_{1,5}-$4$_{1,4}$ E  & -4.4871  & 11  & 15.82    \\ 
95947.4413   & 5$_{0,5}-$4$_{0,4}$ E  & -4.4444  & 11  & 13.93  \\ 
95963.4618   & 5$_{0,5}-$4$_{0,4}$ A  & -4.4441  & 11  & 13.84    \\ 
114940.1789  & 6$_{0,6}-$5$_{0,5}$ E  & -4.2159  & 13  & 19.46    \\ 
114959.9048  & 6$_{0,6}-$5$_{0,5}$ A  & -4.2156  & 13  & 19.36    \\ 
133830.4954  & 7$_{0,7}-$6$_{0,6}$ E  & -4.0256  & 15  & 25.87   \\ 
133854.1003  & 7$_{0,7}-$6$_{0,6}$ A  & -4.0253  & 15  & 25.78   \\ 
138285.0010  & 7$_{1,6}-$6$_{1,5}$ E  & -4.0094  & 15  & 28.92  \\ 
138319.6276  & 7$_{1,6}-$6$_{1,5}$ A  & -4.0100  & 15  & 28.85  \\    
149505.1314  & 8$_{1,8}-$7$_{1,7}$ E  & -3.8898  & 17  & 34.67  \\    
149507.4671  & 8$_{1,8}-$7$_{1,7}$ A  & -3.8897  & 17  & 34.59  \\    
157937.7016  & 8$_{1,7}-$7$_{1,6}$ E  & -3.8445  & 17  & 36.50  \\ 
157974.5872  & 8$_{1,7}-$7$_{1,6}$ A  & -3.8442  & 17  & 36.43  \\ 
168088.6224  & 9$_{1,9}-$8$_{1,8}$ E  & -3.7465  & 19  & 42.74 \\ 
168093.4506  & 9$_{1,9}-$8$_{1,8}$ A  & -3.7464  & 19  & 42.66 \\ 
171265.4737  & 9$_{0,9}-$8$_{0,8}$ E  & -3.7236  & 19  & 41.42  \\  
171296.9876  & 9$_{0,9}-$8$_{0,8}$ A  & -3.7233  & 19  & 41.33  \\

\hline 
\end{tabular}
\label{tab:ch3cho}
\end{table}

\end{appendix}

\end{document}